\documentclass[fleqn,usenatbib]{mnras} % ,draft]

\usepackage{newtxtext}
\usepackage{newtxmath}

\usepackage[T1]{fontenc}

\DeclareRobustCommand{\VAN}[3]{#2}
\let\VANthebibliography\thebibliography
\def\thebibliography{\DeclareRobustCommand{\VAN}[3]{##3}\VANthebibliography}

\usepackage{graphicx}	% Including figure files
\usepackage{amsmath}	% Advanced maths commands
\usepackage{amssymb}	% Extra maths symbols
\usepackage{subfigure}
\usepackage{float,lscape}

\title[G4Jy: Multiwavelength data and analysis]{The GLEAM 4-Jy (G4Jy) Sample: IV. Multiwavelength data and analysis}

\author[White et al. (2026b)]{Sarah V. White$^{1,2,3}$\thanks{sarahwhite.astro@gmail.com}, Precious K. Sejake$^{4}$, Kshitij Thorat$^{4}$, Heinz Andernach$^{5,6}$, Thomas M.O. Franzen$^{7}$,
\newauthor O. Ivy Wong$^{8,9}$, Anna D. Kapi{\'n}ska$^{10}$, Joseph R. Callingham$^{11,12}$, Christopher J. Riseley$^{13,14}$,   \newauthor Nick Seymour$^{3}$, Randall Wayth$^{15}$, Lister Staveley--Smith$^{9}$, Rajan Chhetri$^{3}$, Natasha Hurley-Walker$^{3}$,  \newauthor John Morgan$^{3}$, Paul Hancock$^{3}$, Francesco Massaro$^{16,17,18}$,
 Abigail Garc{\' i}a--P{\' e}rez$^{16,19,20}$, 
\newauthor Ana Jim{\' e}nez--Gallardo$^{20}$,  and  Harold A. Pe{\~ n}a--Herazo$^{21}$  \\
$^{1}$South African Astronomical Observatory (SAAO), PO Box 9, Observatory, 7935, South Africa\\
$^{2}$Department of Physics and Electronics, Rhodes University, PO Box 94, Grahamstown, 6140, South Africa\\
$^{3}$International Centre for Radio Astronomy Research (ICRAR), Curtin University,  Bentley, WA 6102, Australia\\
$^{4}$Department of Physics, University of Pretoria, Hatfield, Pretoria, 0028, South Africa\\
$^{5}$Th{\" u}ringer Landessternwarte, Sternwarte 5, D-07778 Tautenburg, Germany \\
$^{6}$Permanent address: Depto. de Astronom{\' i}a, DCNE, Univ. de Guanajuato, Callej{\' o}n de Jalisco s/n, C.P. 36023 Guanajuato, Mexico\\
$^{7}$Square Kilometre Array Observatory (SKAO), Jodrell Bank, Lower Withington, Macclesfield, SK11 9FT, UK\\
$^{8}$CSIRO Space \& Astronomy, PO Box 1130, Bentley, WA 6102, Australia\\
$^{9}$ICRAR, University of Western Australia (M468), 35 Stirling Highway, Crawley, WA 6009, Australia\\
$^{10}$National Radio Astronomy Observatory (NRAO), 1003 Lopezville Rd, Socorro NM 87801, USA\\
$^{11}$Netherlands Institute for Radio Astronomy (ASTRON), Oude Hoogeveensedijk 4, 7991 PD, Dwingeloo, The Netherlands\\
$^{12}$Anton Pannekoek Institute for Astronomy, University of Amsterdam, Science Park 904, 1098 XH, Amsterdam, The Netherlands\\
$^{13}$Astronomisches Institut, Ruhr-Universität Bochum
Universit{\" a}tsstra$\beta$e 150, 44801 Bochum, Germany\\
$^{14}$Ruhr Astroparticle and Plasma Physics Center (RAPP Center), 44780 Bochum, Germany \\
$^{15}$SKAO, 26 Dick Perry Ave, Kensington, WA 6151, Australia \\
$^{16}$Dipartimento di Fisica, Universit{\` a} degli Studi di Torino, via Pietro Giuria 1, I-10125 Torino, Italy\\
$^{17}$Istituto Nazionale di Astrofisica (INAF) - Osservatorio Astrofisico di Torino, via Osservatorio 20, 10025 Pino Torinese, Italy\\
$^{18}$Istituto Nazionale di Fisica Nucleare (INFN) - Sezione di Torino, via Pietro Giuria 1, I-10125 Torino, Italy\\
$^{19}$Instituto Nacional de Astrof{\' i}sica, {\' O}ptica y Electr{\' o}nica, Luis Enrique Erro 1, Tonantzintla, Puebla 72840, M{\' e}xico\\
$^{20}$European Southern Observatory (ESO), Alonso de C{\' o}rdova 3107, Vitacura, Regi{\' o}n Metropolitana, Chile\\
$^{21}$East Asian Observatory (EAO), 660 N. A’oh{\= o}k{\= u} Place, Hilo, HI 96720, USA\\
}

\date{Accepted 2026 March 4. Received 2026 February 7; in original form 2025 August 8}

\pubyear{2015}

\begin{document}
\label{firstpage}
\pagerange{\pageref{firstpage}--\pageref{lastpage}}
\maketitle

\begin{abstract} % KEEP IT SIMPLE
We provide an updated `multiwavelength' version of the G4Jy catalogue (available at https://github.com/svw26/G4Jy, https://zenodo.org/communities/g4jy/records, and through VizieR), which has 127 new host-galaxy identifications, as described in Paper III of this paper series. We also supplement the redshift information ($0.0 < z < 3.6$), gathered in Paper III,  with $griz$ photometry available through DR10 of the DESI Legacy Surveys. Together, this legacy dataset allows us to investigate the multiwavelength properties of these southern radio-bright galaxies, which includes an initial analysis of radio spectral-curvature for this complete sample ($S_{\mathrm{151\,MHz}} > 4$\,Jy). For example, we present (for the first time in the literature) the radio-power--size diagram as a function of radio spectral-curvature, [$P$--$D$](SCI), noting that the spectral-curvature index (SCI) can act as a proxy for the spectral age of the radio source. This radio-power--size--age diagram shows a predominance of radio galaxies with SCI $>0.15$ and $D < 200$\,kpc, which are candidates for both remnant radio-galaxies and young radio-sources, and a vast range of linear sizes for candidate restarted radio-galaxies (having SCI $<-0.15$). We also show that (i) G4Jy sources populate the entirety of {\it WISE} colour-colour space, (ii) optically point-like sources (i.e. candidate quasars) are brighter than the well-studied $K$--$z$ relation (as expected), and (iii) there is no relation between the SCI of the radio source and its host-galaxy properties.     
\end{abstract}


%%%%%%%%%%%%%%%%%%%%%%%%%%%%%%%%%%%%%%%%%%%%%%%%%%

%%%%%%%%%%%%%%%%% BODY OF PAPER %%%%%%%%%%%%%%%%%%

\section{Introduction}

It is now well-established that at the centre of every galaxy is a supermassive black-hole \citep[e.g.][]{Ryle1967,Richstone1998,EHT2019,Do2019}, with galaxies forming and growing in a hierarchical fashion \citep[e.g.][]{Toomre1972,RoccaVolmerange1990,White1991,Kauffmann1998}. If there is accretion of material onto the black hole, resulting in emission that supplements that of the host galaxy, we describe the galaxy as being `active' and the central region as being an `active galactic nucleus' \citep[AGN; see the review by ][and references therein]{Antonucci1993}. 

AGN are assumed to be axisymmetric and, as such, their appearance (i.e. observed properties) depend upon the angle with which we view them \citep[e.g.][]{Lawrence1982,Urry1995}. Therefore, it is vital that we study them at multiple wavelengths, in order to build up a more-cohesive picture regarding the physical processes that take place in these exciting sources. For example, {\color{black}relevant for radiatively-efficient AGN,} the accretion disc is visible in the optical when viewing the AGN $\sim$face-on, and gives rise to the broad-line and narrow-line regions through photoionisation \citep[e.g.][]{Pogge1988}. However, the putative dusty `torus' \citep[e.g.][]{Pier1992} that surrounds the accretion disc will block the innermost optical light if the source is viewed edge-on. Meanwhile, that same  warm/hot dust re-radiates in the mid-infrared, allowing both dust-obscured and dust-unobscured AGN to be detected \citep[][]{Wright2010,Stern2012}. Similarly, the far-infrared allows us to probe the cooler dust that is heated by star-formation processes in the host galaxy \citep[e.g.][]{Smith2012}, and so is direction-independent \citep[e.g.][]{Sanders1989}. The downside is that far-infrared observations are biased towards high star-formation rates, with dust mass being degenerate with dust temperature \citep[][]{Dunne2000,Dunne2001,daCunha2010}. 

% https://ned.ipac.caltech.edu/level5/Sept14/Conroy/Conroy6.html

Ultimately, we want to understand how physical processes affect the evolution of galaxies. This could be the way in which AGN activity affects star formation, and so either promoting \citep[e.g.][]{Ishibashi2012, Silk2013} or suppressing \citep[e.g.][]{Croton2006, Davies2020, Lammers2023} the growth of the host galaxy. {\color{black}A key} advantage of radio observations is that they are sensitive to {\it both} AGN activity and star-formation processes \citep[e.g.][]{White2015,White2017,White2023}, independent of the amount of dust along the line-of-sight \citep[e.g.][]{DeBreuck2000,Collier2014, Singh2014}. They therefore give us {\color{black}a more-complete} census of galaxy-evolution processes across cosmic time \citep[e.g.][]{Madau2014,McAlpine2015}.  

However, studies of the radio properties of AGN are typically conducted at mid/high radio-frequencies ($\gtrsim1$\,GHz), where radio cores, jets, and hotspots are subject to relativistic beaming \citep{Rees1966}. This leads to a bias in the orientation of the jet axis with respect to the line-of-sight \citep[e.g.][]{Lister2003}, and likely impacts upon the conclusions that we draw from population studies. Like the revised Third Cambridge Catalogue of Radio Sources \citep[3CRR;][]{Laing1983}, the G4Jy Sample \citep{White2018,White2020a,White2020b} does not suffer from this bias, due to its selection at low radio-frequencies ($S_{\mathrm{151\,MHz}} > 4$\,Jy). This is because low-frequency radio emission is dominated by the radio lobes \citep[see the review by][]{Hardcastle2020}, which do not exhibit beaming effects and therefore provide a more `isotropic' view of AGN \citep{Barthel1989}. Furthermore, the fact that low radio-frequencies allow the cumulative effect of AGN activity to be probed over longer timescales \citep[e.g.][]{Kardashev1962,Pacholczyk1970,Jaffe1973} means that we can build a better understanding of the variety in AGN life-cycles \citep[e.g.][]{Harwood2013,Turner2015,Shabala2020} compared with studies restricted to mid/high frequencies. This is because the latter are biased towards to more-recent AGN (or star-formation) activity.  

However, the radio continuum does not have characteristic spectral features that allow the redshift of the source to be derived \citep[e.g.][]{Condon1992}, and so it is vital to combine radio-continuum datasets with information at other wavelengths (or radio spectral-line information, where possible) in order to draw scientific conclusions \citep[e.g.][]{Mauch2007,Best2012,Franzen2021}. For example, \citet{Sabater2019} and \citet{Dabhade2020} have combined the first data release of the LOFAR Two-metre Sky Survey \citep[LoTSS DR1;][]{Shimwell2019, Williams2019} with optical spectroscopy from the Sloan Digital Sky Survey \citep[SDSS;][]{York2000,Abazajian2009} in order to study local radio-detected AGN and identify giant radio-galaxies (GRGs), respectively. \citet{Sabater2019} find that the most massive galaxies have AGN that are always `switched on', and \citet{Dabhade2020} show that their sample of 239 GRGs have spectral indices like those of `normal-sized' radio-galaxies. Both results prompt us to more-carefully consider the `duty cycle' of the AGN, and how these sources interact with their environment \citep[e.g.][]{Hardcastle2013,Hardcastle2014}.  

We know that AGN must exhibit a `duty cycle' (which is the ratio of the time it spends in an `on' state to the time it spends in an `off'/quiescent state) because several sources show restarted AGN activity. For example, \citet{Schoenmakers2000} identified `double-double' radio galaxies where the outer pair of radio lobes is the result of old AGN activity, whilst the inner pair of radio lobes is the result of more-recent activity. A more-unusual case of recurrent activity is that of G4Jy 1080 (IC 4296), where past activity has carved out a pair of tunnels in the surrounding medium, which are then filled with plasma that reaches equilibrium \citep{Condon2021}. At a later time, a new set of jets appear in the same direction as the tunnels, as evidenced via the MeerKAT intensity-map and accompanying spectral-index map (noting that flatter radio emission indicates compact, ongoing activity associated with e.g. the radio core). 

Equally important are so-called `remnant radio-galaxies', where the radio core is {\it not} detected and the only sign of AGN activity is the past emission in the form of radio lobes \citep[e.g.][]{Murgia2011, Mahatma2018, Quici2021}. This importance is because, as shown by theoretical models \citep[e.g.][]{Turner2018,Turner2023}, a complete sample is needed in order to understand the relative proportions of different types of radio galaxy, and so constrain the typical lifecycle for the full population \citep[e.g.][]{Shabala2020}. These models are aided by spectral-curvature information \citep[as suggested by][]{Turneretal2018}, and so, clearly, a multifrequency view is required to understand the entire lifecycle of AGN.

Thankfully, help is at hand with the excellent spectral coverage that is provided by the Murchison Widefield Array \citep[MWA;][]{Tingay2013}, through the GaLactic and Extragalactic All-sky MWA \citep[GLEAM;][]{Wayth2015} Survey (i.e. 20 flux-density measurements at 72 to 231 MHz). The G4Jy Sample was constructed from this low-frequency survey, and will allow models of powerful AGN to be tested more robustly than previously \citep[e.g.][]{Mullin2008,Wang2008,Turner2015}.

% tendency of `radio-loud' AGN to reside in dense environments \citep{Venemans2007,Wylezalek2013, Hlavacek-Larrondo2015}.  However, when studying the properties of powerful AGN as a function of redshift and/or environment, detailed research is hindered by small-number statistics.
 
%Currently, the revised Third Cambridge Catalogue of Radio Sources \citep[3CRR;][]{Laing1983} is the most-prominent, low-frequency radio-source sample ($S_{\mathrm{178\,MHz}} > 10.9$\,Jy) that is optically complete, but this consists of only 173 sources (all in the northern hemisphere).

\subsection{Paper outline} 

This is the fourth paper in the G4Jy paper series, and is outlined as follows: Section~\ref{sec:data} describes the multiwavelength datasets that we consider for our multiwavelength analysis, further to the redshift information collated for Paper III. This is followed by additional detailing of the multiwavelength G4Jy catalogue in Section~\ref{sec:catalogue}. We then present and discuss the results of studying the radio, mid-infrared, and near-infrared properties of G4Jy sources in Section~\ref{sec:results}. Our conclusions are described in Section~\ref{sec:conclusions}, the G4Jy-catalogue metadata are provided in Appendix~\ref{app:catalogue}, additional figures are presented in Appendix~\ref{app:alpha}, and giant radio-galaxies are discussed in Appendix~\ref{sec:GRGs}. 

J2000 co-ordinates and AB magnitudes are used throughout this work, except where {\it WISE} magnitudes are considered (e.g. Section~\ref{sec:MIR_data}) -- these are in the Vega system. We apply a $\Lambda$CDM cosmology, with $H_{0} = 70$\,km\,s$^{-1}$\,Mpc$^{-1}$, $\Omega_{m}=0.3$, $\Omega_{\Lambda}=0.7$.

%Simple mathematics can be inserted into the flow of the text e.g. $2\times3=6$
%or $v=220$\,km\,s$^{-1}$, but more complicated expressions should be entered as a numbered equation:
%\begin{equation}
%    x=\frac{-b$\pm$\sqrt{b^2-4ac}}{2a}.
%	\label{eq:quadratic}
%\end{equation}
%Refer back to them as e.g. equation~(\ref{eq:quadratic}).

%Figures are referred to as e.g. Fig.~\ref{fig:example_figure}, and tables as
%e.g. Table~\ref{tab:example_table}.

%\section{Data}

%In this section we describe the deep, multiwavelength datasets that are used for the selection and analysis of sources over two fields:
%\begin{enumerate}
 %   \item The Cosmic Evolution Survey region (COSMOS; \citealt{Scoville2007}), centred at 10:00:28.6, +02:12:21.0 (J2000) and covering 1.6 sq deg. 
%    \item The XMM-{\it Newton} Large Scale Structure (XMM-LSS) field \citep{Pierre2004}. This is centred at 02:18:00.0, $-$7:00:00.0 (J2000), and we consider the area over which deep near-infrared data is available, which covers 3.5 sq deg. 
%\end{enumerate}
% Where the positions came from:
% https://cosmos.astro.caltech.edu/page/astronomers
% https://vela.astro.uliege.be/themes/spatial/xmm/LSS/srv_char_e.html

\section{Data}
\label{sec:data}

Whilst work towards the `multiwavelength' G4Jy catalogue is based on numerous datasets, in this section we only describe those that are of direct relevance for the multiwavelength analysis that is presented in this paper. For example, the newest dataset that we consider is the tenth data release of the DESI Legacy (Imaging) Surveys \citep{Dey2019}, which is summarised in Section~\ref{sec:optical_data}. For further details of the radio data, we refer the reader to Papers I, II, and III.

\subsection{Radio data}
\label{sec:radio_data}

The multi-frequency coverage provided by the MWA \citep{Tingay2013} enables excellent low-frequency constraints of radio emission, with 20 flux-density measurements (from 72\,MHz to 231\,MHz) provided through the GLEAM Survey \citep{Wayth2015}. The 151-MHz sub-band flux-density was used to search for, and select, radio sources with a total, integrated $S_{\mathrm{151\,MHz}}$ greater than 4\,Jy, thereby creating the G4Jy Sample \citep{White2020a,White2020b}. These radio sources are distributed across the southern sky (Dec. $<+30$\,deg, $|b| > 10$\,deg), following the footprint of the GLEAM Extragalactic Catalogue \citep{HurleyWalker2017}. The details of individual GLEAM components are provided in the original G4Jy catalogue \citep{White2020a,White2020b} -- including the GLEAM components that required re-fitting -- whilst the {\it total}, integrated flux-densities for the 20 sub-bands are retained for the updated `multiwavelength' G4Jy catalogue. 

The two G4Jy catalogues can easily be combined via cross-matching on the radio-centroid positions. These were calculated by \citet{White2020a,White2020b} via the imaging and catalogues provided by (i) the Sydney University Molonglo Sky Survey \citep[SUMSS;][]{Mauch2003, Murphy2007} at 843\,MHz, and (ii) the NRAO (National Radio Astronomy Observatory) VLA (Very Large Array) Sky Survey \citep[NVSS;][]{Condon1998} at 1.4\,GHz. Both surveys have a spatial resolution of $\sim$45\,arcsec, which allowed \citet{White2020a,White2020b} to (a) visually assess the radio morphology of G4Jy sources (`single', `double, `triple', `complex'), and (b) estimate their angular sizes, as consistently as possible.

\citet{White2020a,White2020b} also recorded the total flux-densities measured via SUMSS- and NVSS-catalogue data (see section 5.2 of Paper III), and from these measurements, they derived two-point spectral indices (G4Jy\_SUMSS\_alpha = $\alpha_{\mathrm{151\,MHz}}^{\mathrm{843\,MHz}}$, and G4Jy\_NVSS\_alpha = $\alpha_{\mathrm{151\,MHz}}^{\mathrm{1400\,MHz}}$), following an assumed power-law description of the radio emission ($S_{\nu} \propto \nu^{\alpha}$). Such an assumption was also made when calculating the 20-point, low-frequency spectral-index, G4Jy\_alpha = $\alpha_{\mathrm{72\,MHz}}^{\mathrm{231\,MHz}}$ (based upon the total, integrated flux-densities provided in the G4Jy catalogue). For reference, the 5-$\sigma$ surface-brightness limits of the relevant surveys (GLEAM, SUMSS, and NVSS) are $S_{\mathrm{200\,MHz}} \gtrsim 50$\,mJy\,beam$^{-1}$, $S_{\mathrm{843\,MHz}} \simeq$\,6--10\,mJy\,beam$^{-1}$, and $S_{\mathrm{1400\,MHz}} \simeq 2.5$\,mJy\,beam$^{-1}$, respectively.

For the current work, we also refer to MeerKAT and VLASS (VLA Sky Survey; \citealt{Lacy2020}) radio-images, which are made available to the community by \citet{Sejake2023}, \citet[][]{Gordon2020}, and Dong et al., in prep. These images have a spatial resolution of $\sim$7\,arcsec and $\sim$2.5\,arcsec, respectively, and we use them to obtain more-accurate angular-sizes for bonafide giant radio-galaxies (Table~\ref{tab:GRGs}). However, as in Paper III, we note that updating the angular sizes in the G4Jy catalogue is deferred to future work.

\subsection{Mid-infrared data}
\label{sec:MIR_data}

In Papers II \citep{White2020a} and III \citep{White2026}, we (predominantly) identified the host galaxy of G4Jy sources in AllWISE-W1 mid-infrared images \citep{Cutri2013} in order to avoid the bias incurred by dust-obscuration (which affects optical surveys). This band is at 3.4\,$\upmu$m, and is accompanied by bands W2 (4.6\,$\upmu$m), W3 (12\,$\upmu$m), and W4 (22\,$\upmu$m), courtesy of the {\it Wide-field Infrared Survey Explorer} ({\it WISE}; \citealt{Wright2010}). The all-sky survey conducted with {\it WISE} resulted in 5-$\sigma$ sensitivities of 0.054, 0.071, 0.73, and 5.0\,mJy, respectively, at the following corresponding resolutions: 6.1, 6.4, 6.5, and 12.0\,arcsec. In terms of signal-to-noise ratios (SNRs), we find that six sources in the G4Jy Sample have SNR < 5 for the W1 band. These are G4Jy 573, G4Jy 764, G4Jy 1299, G4Jy 1312, G4Jy 1341, and G4Jy 1576, due to being either very mid-infrared faint or affected by nearby bright mid-infrared emission. In addition, we quote the AllWISE measurements as (Vega) magnitudes, noting that the $>$95-per-cent completeness levels are 17.1, 15.7, 11.5, and 7.7\,mag in W1, W2, W3 and W4, respectively \citep{Cutri2013}\footnote{\url{https://wise2.ipac.caltech.edu/docs/release/allwise/expsup/sec2_1.html}}. 
% G4Jy 573 and G4Jy 1576 are the faint ones

\subsection{Near-infrared data}
\label{sec:NIR_data}

Magnitudes are presented in the G4Jy catalogue for the near-infrared bands, $J$, $H$, and $K_{\mathrm{s}}$, from the Two Micron All-Sky Survey (2MASS; \citealt{Cutri2003,Skrutskie2006}) via the AllWISE catalogue. These observations are from two 1.3-m telescopes, located at Cerro Tololo in Chile and Mount Hopkins in Arizona, and dedicated to the survey. The result is $\sim$all-sky coverage to the following 10-$\sigma$ depths, with apertures of 4-arcsec radius: $J=15.8$, $H=15.1$, and $K_{\mathrm{s}}=14.3$\,mag (in the Vega-magnitude system). The resolution of the imaging is 2.5\,arcsec -- this being the result of 2\,arcsec\,pixel$^{-1}$ detectors, the instrument response, and the best seeing conditions -- and the astrometric accuracy is of the order 100\,mas. 2MASS constructed a Point-Source Catalogue of 470,992,970 sources, and an Extended Source Catalogue \citep{Jarrett2000} containing 1,647,599 sources. G4Jy sources that have their host galaxy identified in the 2MASS Extended Source Catalogue are denoted with the prefix of `2MASX'.  

%As part of future work, we plan to update the G4Jy catalogue with $YJHK_{\mathrm{s}}$ photometry from the VISTA (Visible and Infrared Survey Telescope for Astronomy) Hemisphere Survey \citep[VHS;][]{McMahon2013, McMahon2021}, accompanied by visual inspection of $K_{\mathrm{s}}$-band overlays for the full sample. For now (e.g. Section~\ref{sec:new_IDs}), we note that the 5-$\sigma$ point-source limiting magnitudes for VHS are: $Y = 21.1, J=20.8, H=20.6$, and $K_{\mathrm{s}}=20.0$\,mag (in the AB-magnitude system).

\subsection{Optical photometry}
\label{sec:optical_data}

We turn to Data Release 10 (DR10) of the DESI Legacy (Imaging) Surveys \citep{Dey2019} for optical photometry in the $g$, $r$, $i$ and $z$ bands. These data have been obtained through the Beijing--Arizona Sky Survey (BASS; \citealt{Zou2017}) and the Mayall $z$-band Legacy Survey (MzLS; \citealt{Silva2016}) for sources above a Declination of 32.375\,deg, and through the Dark Energy Camera (DECam; \citealt{Flaugher2015}) Legacy Survey (DECaLS) for sources below this Declination\footnote{A reminder that the entirety of the G4Jy Sample is at Dec. $<30.0$\,deg.}. For homogeneity, {\sc legacypipe}\footnote{\url{https://github.com/legacysurvey/legacypipe}} software has been run to perform all of the steps from calibrating images to creating catalogues. It includes the running of {\sc SourceExtractor} \citep{Bertin1996} and {\sc PSFEx} \citep{Bertin2011} to generate light-profile models, which are then passed to an algorithm called {\sc The Tractor} \citep{Lang2016} for model-fitting and photometry. The 5-$\sigma$ AB-magnitude depths reached are $g=24.0$, $r=23.4$, and $z=22.5$\,mag \citep{Dey2019}, and the 50th-percentile depths\footnote{These values were made available through the Astro Data Lab User Forum, \url{https://datalab.noirlab.edu/help/} -- thank you!} are $g=24.9$, $r=24.4$, $i=24.0$, and $z=23.4$\,mag. Meanwhile, the astrometric precision is better than $\pm$0.03\,arcsec. This is having used the epoch of a given Legacy Surveys observation to predict {\it GAIA} (Global Astrometric Interferometer for Astrophysics; \citealt{Lindegren1994}) positions to which certain source positions are tied. In doing so, the source extraction becomes fixed to the {\it GAIA} Data Release 2 \citep{Gaia2018} system. For details of the information that is retained for the G4Jy catalogue, see Section~\ref{sec:xmatch_photometry}.

\subsection{Optical spectroscopy}
\label{sec:optical_spectroscopy}

In Paper III we described the multiple optical-spectroscopy datasets that we have gathered, assessed, and re-fitted (where necessary) for the G4Jy Sample, with the aim of reaching 100 per cent spectroscopic completeness. This is crucial for taking full advantage of the lack of bias in the low-frequency selection of the sources, and so allowing the most-robust comparisons with theoretical models. We list these spectroscopy datasets in table 2 of \citet{White2026} and, as a result of this thorough work, the G4Jy Sample currently has a spectroscopic completeness of 34 per cent and a redshift range of $0.00075 < z < 3.56990$.  %For interest, we present the six highest-redshift sources ($z > 2.75$) in Figure~\ref{fig:highest_zs}.  % Just point-sources, so not learning anything from this figure

\section{The G4Jy multiwavelength catalogue}
\label{sec:catalogue}

The original G4Jy catalogue provided detailed information for the low-frequency radio emission of all 1,863 G4Jy sources, as well  as mid-infrared data through 1,606 host-galaxy identifications in AllWISE (which itself is accompanied by near-infrared data from 2MASS). Thanks to supplementary work on identifying the host galaxies of the radio emission (described in Paper III), we now have a total of 1,733 identifications for collating additional multiwavelength information (this section) and conducting further analysis (Section~\ref{sec:results}). 
% 1863 - 128 - 2'n' = 1733

\subsection{Crossmatching photometry}
\label{sec:xmatch_photometry}

We obtain optical photometry (Section~\ref{sec:optical_data}) through a crossmatch of the G4Jy catalogue with the Legacy Surveys DR10 \citep{Dey2019}. This is performed on the host-galaxy positions (given by host\_RAJ2000, host\_DEJ2000) with a radius of 1\,arcsec, to be conservative. This is because 1\,arcsec is the typical resolution of the VISTA Hemisphere Survey DR5 \citep{McMahon2021} imaging, within which 17 of the G4Jy sources were identified. (Note that the prefix of the host\_name gives an indication of the method/data through which a particular source was identified.)   

The Legacy Surveys DR10 provides a wealth of information for the southern sky\footnote{\url{https://www.legacysurvey.org/dr10/files/\#tractor-catalogs-south-tractor}}, and the columns that are retained for our purposes are listed in Table~\ref{tab:catalogue_columns} (Appendix~\ref{app:catalogue}). This includes the optical morphological type\footnote{\url{https://www.legacysurvey.org/dr10/description/\#morphological-classification}} (labelled `morphtype\_LSDR10' in the G4Jy catalogue; see Table~\ref{tab:sersic_index}), and we direct the reader to Section~\ref{sec:WISE_diagrams} for how we use this in combination with {\it WISE} colour-colour space. In addition to the $griz$ magnitudes are `de-reddened' magnitudes for these optical bands, which have been determined through fitting for Galactic extinction. We also retain information regarding the signal-to-noise per optical band, the size of the point-spread function (PSF), and the S{\' e}rsic index that is determined alongside the optical morphology (Table~\ref{tab:sersic_index}).

\begin{table}
\centering 
\caption{The optical morphology types that are fitted by {\sc The Tractor} \citep{Lang2016} as part of photometric-catalogue creation for the DESI Legacy Surveys \citep{Dey2019}. The `Fitting' column indicates whether the S{\' e}rsic index for that particular morphology type is fixed or is allowed to vary, and `0' is a placeholder value in the `S{\' e}rsic index' column for `PSF' sources. These results are crossmatched into the G4Jy catalogue (Appendix~\ref{app:catalogue}) via a matching radius of 1\,arcsec on the host-galaxy positions (Section~\ref{sec:xmatch_photometry}).}
\begin{tabular}{@{}lccc@{}}
 \hline
Morph. & Description of the  & S{\' e}rsic & Fitting \\
type & optical light profile  & index &  \\
 \hline
PSF & Point-spread function (for point-sources) & 0 & fixed  \\
EXP & Exponential disc (for spiral galaxies) & 1 & fixed  \\
REX & Round exponential of variable radius & 1 & fixed  \\
DEV & \citet{deVaucouleurs1948} $r^{-1/4}$ power-law & 4 & fixed  \\
SER & \citet{Sersic1963} light profile  & 0.5 to 6.0 & varies \\

\hline
\label{tab:sersic_index}
\end{tabular}
\end{table}

We note that G4Jy names are now fully resolvable in many online databases, and this includes the Legacy Surveys Sky Viewer\footnote{For example, G4Jy~1080 can be viewed via \\ \url{https://www.legacysurvey.org/viewer\#G4Jy\%201080}}. Visual inspection of optical images for the full G4Jy Sample is beyond the scope of the current work, but is planned for the future as part of aiding spectroscopic follow-up (particularly for the optically-fainter sources, at $r>20$\,mag).

%\vspace{-5mm}
\subsection{Redshift information}
\label{sec:z_information}

Within the G4Jy catalogue, spectroscopic redshifts can be found in the `zsp\_misc' column, photometric redshifts can be found in the `zph\_misc' column, and redshifts data-mined via NED\footnote{\url{https://ned.ipac.caltech.edu}} can be found in the `z\_NED' column. In determining a single z\_origin\_flag per source (see table 2 of Paper III), the available redshifts have the following order of preference: zsp\_misc > zph\_misc > z\_NED. Later in this paper we refer to `cumulative redshifts', by which we mean we have a single redshift per source ($z$ = $z_{\mathrm{sp}}$ or $z_{\mathrm{ph}}$ or $z_{\mathrm{NED}}$), following the aforementioned order of redshift preference.

Our spectroscopic-redshift assessment (Paper III) included visually inspecting the available spectroscopy and assigning a `z\_Quality\_flag', Q. A reminder of the meanings of the different quality flags are repeated here for convenience:

\begin{itemize}
\item Q = 1 indicates that the redshift is highly robust, with multiple spectral features detected with good signal-to-noise ratios. This flag is also assigned when a single {\it broad} emission-line\footnote{We caution that a single narrow `line' may be the result of a cosmic ray being detected, or poor background-subtraction near detector-chip edges.} (with a distinctive profile, e.g. Ly-$\alpha$) is detected with very good signal-to-noise.
\item Q = 2 indicates that the redshift is fairly robust, where multiple spectral features are detected but they are less discernible from the continuum or noise level than for Q = 1.
\item Q = 3 indicates that the redshift is a tentative measurement, as suggested by the marginal detection of multiple spectral features against the continuum/noise.
\item Q = 4 indicates that the redshift is unreliable, as it is derived from (a) a single spectral feature detected with low-to-moderate signal-to-noise, (b) one or more spectral features that have been misidentified (resulting in an incorrect redshift calculation), or (c) spectral features that are not discernible by eye.
\item Q = 5 indicates that visual inspection of the spectroscopy has not been completed (e.g. because the spectrum is not available online). This flag is automatically given to all photometric redshifts and all redshifts acquired through automated data-mining of the NASA/IPAC Extragalactic Database (NED). 
\end{itemize}

If there are multiple spectroscopic redshifts available for a given source, the one with the lowest Q value (i.e. the best quality) is retained for the G4Jy catalogue. Our ultimate aim is for 100 per cent of the sample to have spectroscopic redshifts where Q = 1 or 2. 

For interest, we retain the spectroscopic redshifts provided by the \citet{AbdulKarim2025} as a separate column, `zsp\_DESI'. Alongside this is the warning flag\footnote{\url{https://desidatamodel.readthedocs.io/en/latest/bitmasks.html\#zwarn}} (`zsp\_WARN\_DESI'), which is `0' for all DESI-covered sources except: G4Jy 178, G4Jy 227, G4Jy 1165, and G4Jy 1228. For these, the warning flag is `4', which means that there are many outliers in terms of redshift-template fitting. However, following our visual inspection (Paper III), we believe that  zsp\_WARN\_DESI = 4 may be overcautious for G4Jy 227, since we have assigned it a z\_Quality\_flag of 1.

Similarly, we retain the main columns from the photometric catalogues, WISExSuperCOSMOS \citep[wiseScos;][]{Bilicki2016} and DESI LS DR8 photo-z South \citep[LSDR8south;][]{Duncan2022}, for the G4Jy catalogue. Please see Appendix~\ref{app:catalogue} for confirmation of these.

\subsection{Radio data per GLEAM component}

The parent catalogue for the G4Jy Sample is the GLEAM Extragalactic Catalogue \citep{HurleyWalker2017}, which provides detailed characterisation of low-frequency radio-emission on a component-by-component basis (subject to $\sim$2-arcmin spatial resolution via the MWA Phase-I beam). GLEAM components could describe an entire radio-source or an individual radio-lobe (or part thereof), hence the additional work that is needed to create {\it source} catalogues, such as the G4Jy catalogue \citep{White2020a,White2020b}. For the latter we retained information for the 1,960 GLEAM components that correspond to 1,863 G4Jy sources (including 15 components that were re-fitted), and introduced a G4Jy\_component column to enable a short-hand way of referring to a particular GLEAM-component. For example, G4Jy 7A is GLEAM~J000441+124907 and G4Jy 7B is GLEAM J000456+124810, both belonging to the source, G4Jy~7. 

{\bf To simplify the combining of the G4Jy catalogue with other datasets, we hereafter refer to G4Jy sources as opposed to constituent GLEAM components.} (This avoids repeating source properties, such as redshift, across multiple rows for multi-GLEAM-component sources.) As such, only the total, integrated flux-densities (per GLEAM sub-band) and spectral indices are retained for the latest G4Jy catalogue. If the reader would like to combine both the original and the updated G4Jy catalogues, they can do so easily by cross-matching on the radio-centroid positions \citep{White2020a,White2020b}. This will then allow them to access the wealth of low-frequency information that is listed in appendix E of Paper I, such as the local noise-level (Jy\,beam$^{-1}$) in the 72--80\,MHz sub-band image per GLEAM component. \\

\subsection{Ancillary information}

Accompanying the redshifts from NED are: alternative names of G4Jy sources in published work, the fiducial R.A. and Dec. quoted in NED, and the source `type'\footnote{\url{https://ned.ipac.caltech.edu/help/ui/nearposn-list_objecttypes}} (e.g. `RadioS' for radio source) that corresponds to the alternative name. Furthermore, the descriptions for columns that are newly-added to form the G4Jy multiwavlength catalogue can be found in Appendix~\ref{app:catalogue}.

% Don't know why G4Jy 1357 has * = star in the catalogue, as NED gives an alternative name of 4C +30.30 online.

\section{Results and discussion}
\label{sec:results}

In this section we consider the multiwavelength properties of the G4Jy Sample, and incorporate the newly-acquired redshifts \citep[][]{White2025a,White2026} into this analysis. We start with analysis of the spectral curvature that is exhibited through the radio emission at multiple frequencies (Section~\ref{sec:spectral_curvature}), which will be explored further in a future paper (White et al., in prep.). Each of the subsequent sub-sections then connects back to this spectral analysis, which itself is enabled by the excellent low-frequency coverage provided by the MWA, through the GLEAM survey.

\subsection{Radio properties}

\subsubsection{Spectral curvature}
\label{sec:spectral_curvature}

Previously, \citet{White2020b} presented the radio spectra for 67 G4Jy sources that overlap with the 3CRR sample \citep{Laing1983}, G4Jy-3CRR, by way of assessing the flux-density scale (see their table 8 and appendix F). These spectra, like those of \citet{Laing1980}, demonstrate that the bright radio emission from these sources does {\it not} always follow the power-law function ($S_{\nu} \propto \nu^{\alpha}$, where $\alpha$ is the spectral index) that is typically assumed to describe the radio emission. Of course, this is already evident from the presence of peaked-spectrum sources \citep[e.g.][]{Callingham2017}, which are believed to be radio sources in the early stages of their lifetimes. Nevertheless, we often construct two-point spectral indices (i.e. $\alpha$ values that are constrained by flux-densities at two different frequencies) as a simple way of extrapolating measured flux-densities from one frequency to another. This is the purpose intended for the G4Jy\_NVSS\_alpha (G4Jy\_SUMSS\_alpha) values provided in the G4Jy catalogue, which are calculated between 151\,MHz and 1400\,MHz (843\,MHz, respectively) and aid, for example, follow-up observations with MeerKAT \citep{Sejake2023}. 

\begin{figure}
\centering
\subfigure[]{
\includegraphics[scale=0.51]{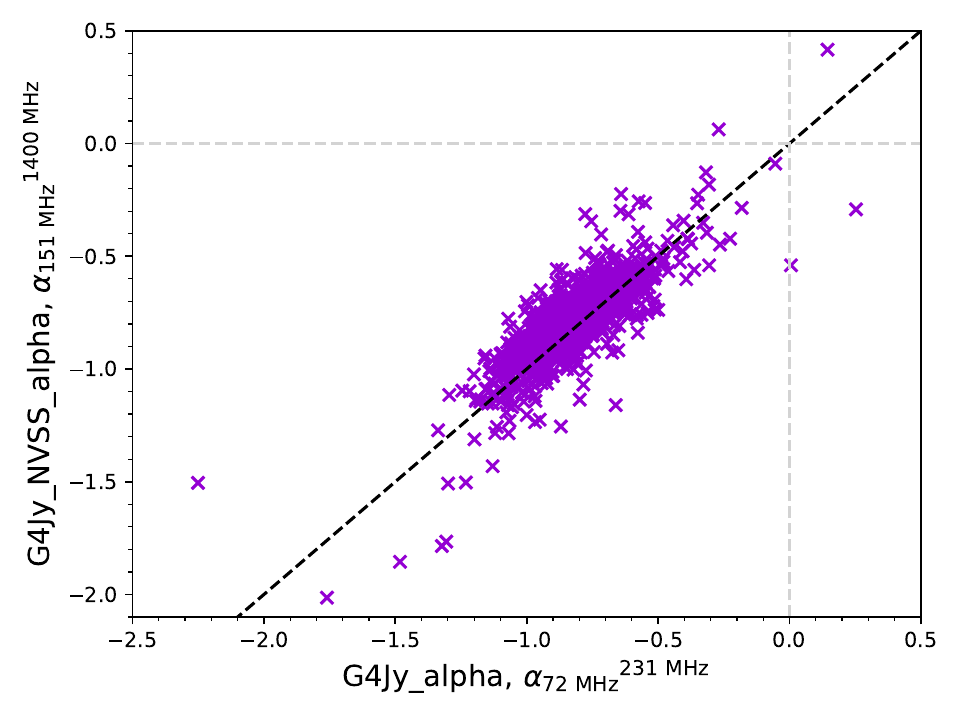} }
\subfigure[]{
\includegraphics[scale=0.51]{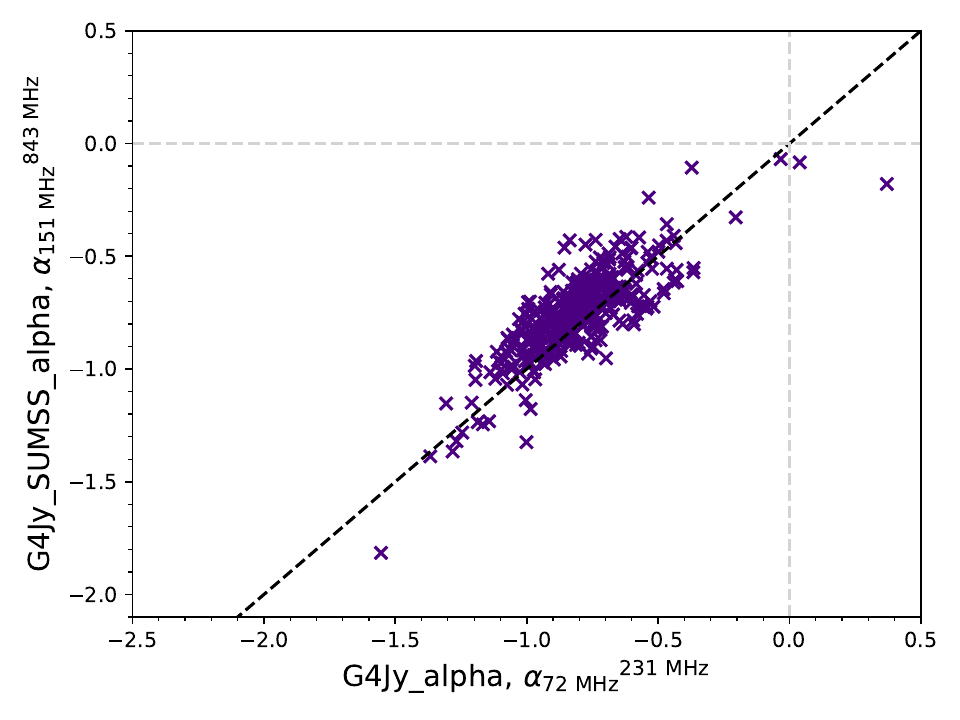} }
\subfigure[]{
\includegraphics[scale=0.51]{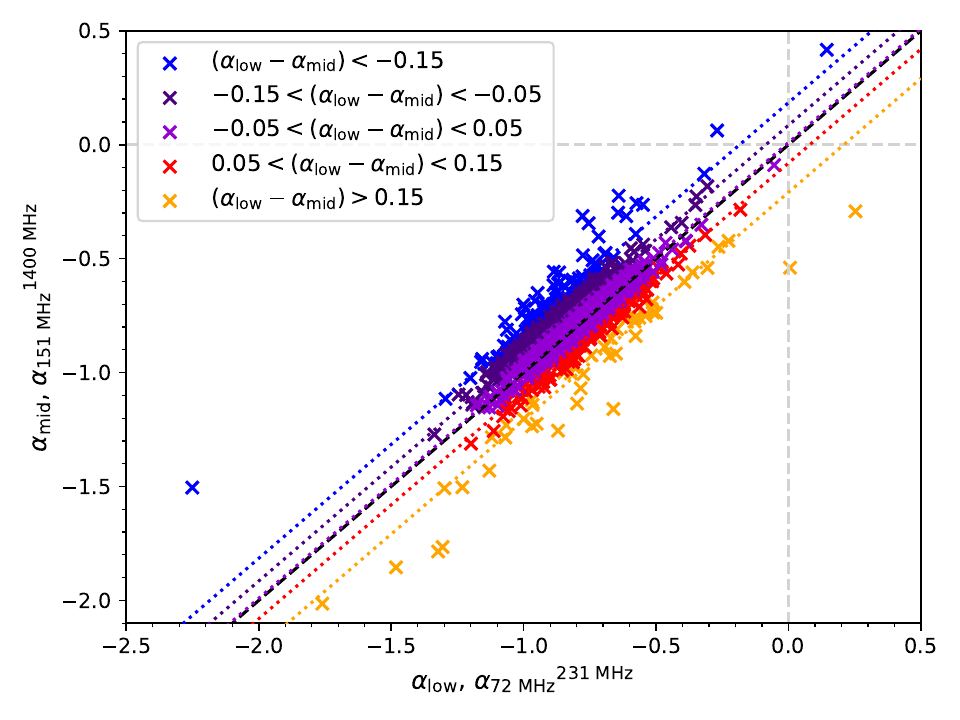} }
\caption{ Distributions in the different spectral indices [G4Jy\_alpha, (a) G4Jy\_NVSS\_alpha, and (b) G4Jy\_SUMSS\_alpha] that are available in the G4Jy catalogue. In each panel, a two-point spectral index (y-axis value) is plotted against the 20-point spectral index derived at low radio-frequencies, G4Jy\_alpha (Section~\ref{sec:spectral_curvature}). The latter is renamed `$\alpha_{\mathrm{low}}$' for panel (c), and each panel has a diagonal dashed-line that represents the 1:1 relation. For panel (c), the NVSS distribution is split into five regions of differing degrees of spectral curvature, as indicated by the $\alpha_{\mathrm{low}} - \alpha_{\mathrm{mid}}$ value in the legend. The diagonal dotted-lines (in blue, indigo, violet, red, and orange) represent the median $\alpha_{\mathrm{low}} - \alpha_{\mathrm{mid}}$ values (i.e. per region), which are provided in Table~\ref{tab:proxy_beta}. }
\label{fig:alpha_alpha}
\end{figure}

Also provided in the G4Jy catalogue is the integrated spectral-index, G4Jy\_alpha, through (power-law) fitting to all 20 flux-densities within the GLEAM band, which spans 72 to 231\,MHz. Accompanied by a reduced-$\chi^2$ value for the fitting, the availability of this spectral index indicates the validity of the low-frequency emission being approximated by $S_{\nu} \propto \nu^{\alpha}$. [Please see \citet{White2020b} for further details.] Therefore, we can assess the degree of spectral curvature {\it beyond} this frequency range by comparing G4Jy\_alpha with G4Jy\_NVSS\_alpha and G4Jy\_SUMSS\_alpha (Figure~\ref{fig:alpha_alpha}ab). If a power-law description remains an accurate approximation, then the two values will equal each other, with the resulting datapoint lying on the 1:1 relation (diagonal dashed-lines) in Figure~\ref{fig:alpha_alpha}. Sources lying above the 1:1 relation have a radio spectrum that can be described as `concave', whilst sources lying below the 1:1 relation have a radio spectrum that can be described as `convex' (with respect to the frequency axis).

Note that there are subtle differences in the distributions of G4Jy\_NVSS\_alpha and G4Jy\_SUMSS\_alpha [see figure 14a of \citet{White2020b}], on account of the higher-frequency end being different (i.e. 1400\,MHz and 843\,MHz, respectively). Coupled with the fact that sources lie at different redshifts, this means that different rest-frame frequencies are being probed. Until we have direct measurements of the whole sample at 1400\,MHz, we only consider the spectral curvature for sources that are at Dec. $>-39.5$\,deg, and are therefore covered by the NVSS footprint. This equates to 77 per cent of the G4Jy Sample, with their spectral indices presented again in Figure~\ref{fig:alpha_alpha}c.

Whilst the vast majority of sources show steep-spectrum emission ($\alpha < -0.5$), the number of sources lying above the 1:1 relation exceeds the number of sources lying below the 1:1 relation. In the case of G4Jy-NVSS sources, this unevenness in the distribution is quantified by 27 per cent of the subsample lying below the line (Figure~\ref{fig:alpha_alpha}a). The corresponding percentage for G4Jy-SUMSS sources is 13 per cent (Figure~\ref{fig:alpha_alpha}b). We also note that sources above the line, where the low-frequency emission is steeper than the mid-frequency emission, could represent examples of restarted radio-galaxies, in agreement with simulations by  \citet{Turner2018}. These simulations also show that the mid-frequency emission is steeper than the low-frequency emission, as one would expect, for ageing radio-galaxies (lying below the 1:1 relation). Therefore, the distribution in spectral curvature allows for constraints to be placed on the lifecycles of a complete AGN sample. %This will be explored further in future work.  

Next, we split the datapoints for NVSS sources (Figure~\ref{fig:alpha_alpha}a) into five regions in $\alpha_{\mathrm{low}}-\alpha_{\mathrm{mid}}$ space (Figure~\ref{fig:alpha_alpha}c), with these categorised spectral-curvature values being of interest for the remainder of the multiwavelength analysis described in this section. Formally, we define the spectral-curvature index (SCI) as:
\begin{equation}
    \mathrm{SCI}_0 = \alpha_{\mathrm{low}} - \alpha_{\mathrm{mid}}.
    \label{eq:sci0}
\end{equation}
Whilst this may appear to be the opposite way round from the definition put forward by \citet{Murgia2011}, the sense in which they define a power-law is also opposite to our definition: $S_\nu \propto \nu^{-\alpha}$. The result is that we have the same physical interpretation of the spectral-curvature-index values, these being SCI$_0 < -0.05$ for candidate restarted/renewed radio-galaxies, and SCI$_0 > 0.05 $ for candidate ageing/remnant radio-galaxies (Figure~\ref{fig:alpha_alpha}d). [We emphasise our use of the word `candidate' to describe these categories, since further analysis (beyond the scope of this work) is needed to determine, e.g., the true fraction of remnant radio-galaxies.] The $-0.05 <$ SCI$_0$ $< 0.05$ interval represents the G4Jy sources that have radio emission that is typically well-described by a power-law function from 72\,MHz to 1400\,MHz, as also indicated by coincidence with the 1:1 relation. For the median spectral-curvature indices (and median spectral-indices) for the five intervals, please see Table~\ref{tab:proxy_beta}, noting that the typical uncertainty in the SCI$_{0}$ value is 0.02. 

  \begin{table}%[]
    \centering
    \begin{tabular}{l|r|c|c}
    \hline
    Spectral curvature &  Median  &  Median 
     & Median \\ 
    index, $\alpha_{\mathrm{low}} - \alpha_{\mathrm{mid}}$ & $\alpha_{\mathrm{low}} - \alpha_{\mathrm{mid}}$ & $\alpha_{\mathrm{low}}$ & $\alpha_{\mathrm{mid}}$
      \\ 
       \hline

      $<-0.15$ &   $-$0.184 & $-$0.887 & $-$0.702 \\
        $-$0.15 to $-0.05$ &  $-$0.085 & $-$0.863 & $-$0.777  \\
$-$0.05 to 0.05 &  $-$0.009  & $-$0.804 & $-$0.798   \\
0.05 to 0.15 &  0.081 & $-$0.767 & $-$0.842  \\
$>0.15$  &   0.209  & $-$0.695 & $-$0.927  \\

    \hline
    \end{tabular}
    \caption{ Median values in the spectral-curvature index, SCI$_0 = \alpha_{\mathrm{low}} - \alpha_{\mathrm{mid}}$  (Section~\ref{sec:spectral_curvature}), for five different regions of Figure~\ref{fig:alpha_alpha}c. For interest, we also provide the median in the low-frequency spectral-index ($\alpha_{\mathrm{low}}$) and the mid-frequency spectral-index ($\alpha_{\mathrm{mid}}$) for these intervals/regions.} 
    \label{tab:proxy_beta}
\end{table}

We note that sources with very old radio-emission, and no signs of recent activity in the radio, should occupy the region defined by SCI$_0>0.15$ (though not exclusively). That is, this is the parameter space within which we expect to find remnant radio-galaxies \citep[][Stewart et al., in prep.]{Murgia2011,Mahatma2018,Quici2021,Quici2025}, which could help (in part) to explain the earlier observation that there are fewer sources in the sample lying below the 1:1 relation. In other words, this may be a selection effect connected to the identification completeness of the G4Jy Sample. The NVSS (and SUMSS flux-densities) are {\it integrated} values, and the spatial distribution of this mid-frequency emission could be such that the radio core is only marginally detected (e.g. G4Jy 1289; \citealt{Sejake2023}). As a result, the  $\alpha_{\mathrm{mid}}$ value will be much steeper than previously estimated, leading to a datapoint that moves towards the lower right-hand corner of the SCI$_0$ =  $\alpha_{\mathrm{low}} - \alpha_{\mathrm{mid}}$ plane. Therefore, obtaining core flux-densities for the full sample, where possible, is the only way in which we can be sure that the remnant radio-galaxy population is properly constrained. [See \citet{Mahatma2018} for further discussion.] 

Before progressing further with the discussion, we caution that the SCI$_0 > 0.05$ sources may include {\it young} radio-galaxies exhibiting a peaked spectrum. (I.e., spectral turnover of the radio emission results in a spectral shape that is flatter at lower frequencies than at higher frequencies.) {\color{black}Such a peaked spectrum is thought to be
due to free-free absorption or synchrotron self-absorption processes
in compact sources}. This complicates the interpretation of the SCI$_0$ value as a proxy for the spectral age of the source, which is why we again emphasise that this category is described as {\it candidate} ageing/remnant radio-galaxies. The ratio of truly-ageing sources to young sources is currently unknown, but will be clarified by ongoing {\color{black}spectral-age analysis} (White et al., in prep.). In the meantime, we note that 10 per cent of SCI$_0 > 0.05$ sources have $\alpha_{\mathrm{low}} < -1.0$ -- a combination that most-clearly can be interpreted as a dying radio-galaxy, and so acts as a lower limit. If we consider the SCI$_0 > 0.15$ subset, this fraction (of NVSS sources with $\alpha_{\mathrm{low}} < -1.0$) increases to 21 per cent.

We also investigate whether there is a trend within the SCI$_0$ = $\alpha_{\mathrm{low}} - \alpha_{\mathrm{mid}}$ plane that is brought to light by a correlation with the spectroscopic redshift (Appendix~\ref{app:alpha}). However, no such trend is visible, which goes against the idea that following-up sources with ultra-steep spectral indices ($\alpha \lesssim -1.3$; \citealt{DeBreuck2000}) is an effective way of identifying high-redshift sources \citep[e.g.][]{Blumenthal1979,Roettgering1994,Saxena2018}. This may be because, as suggested by \citet{GopalKrishna1988}, \citet{Onuora1989} and \citet{Pinjarkar2025}, the apparent connection between spectral index and redshift is rather driven by an underlying correlation between spectral index and radio luminosity (see also Sejake et al., submitted). %(albeit, for lower-luminosity sources than 3CRR and therefore, by extension, the G4Jy Sample). 

% Removed "fully accounted for" because the remanants are still there, it's just that their datapoint may be in the 'wrong' region of alpha-alpha space

\begin{figure}
\centering
\includegraphics[scale=0.52]{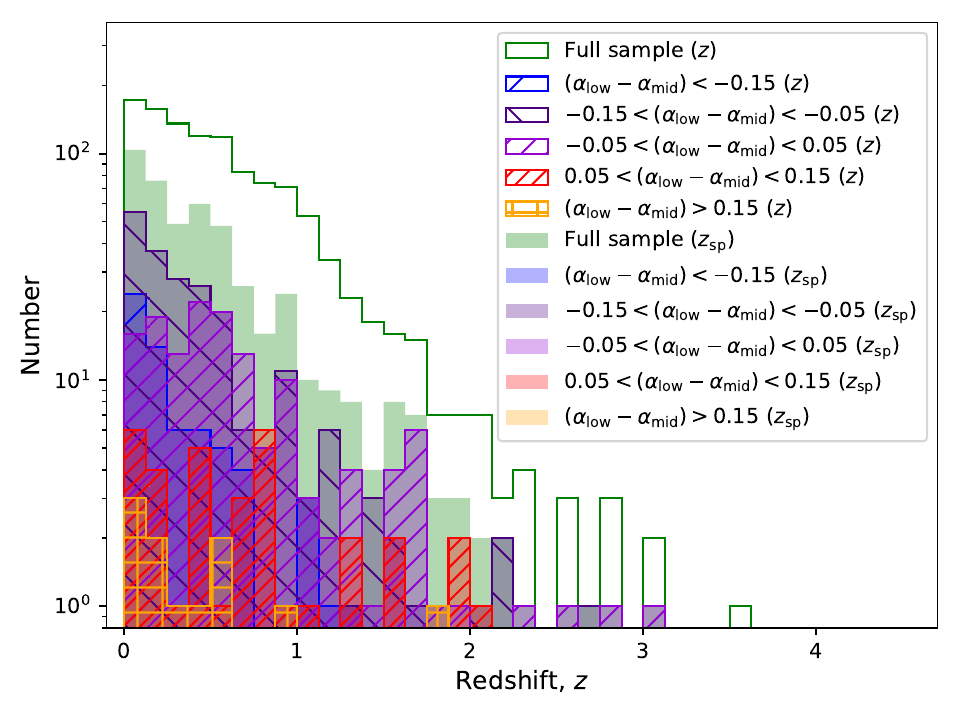} 
\caption{ Distributions of the redshifts for G4Jy-NVSS sources, colour-coded by the degree of spectral curvature (Section~\ref{sec:spectral_curvature}). `Cumulative redshifts' ($z = z_{\mathrm{sp}}$, $z_{\mathrm{ph}}$, $z_{\mathrm{NED}}$) are used to determine the outlined histograms, and spectroscopic redshifts ($z_{\mathrm{sp}}$) are used to determine the shaded histograms. }
\label{fig:redshifts_by_proxybeta}
\end{figure}

In addition, Figure~\ref{fig:redshifts_by_proxybeta} presents the redshift histograms for G4Jy-NVSS sources, as a function of spectral curvature (with spectroscopic redshifts resulting in the shaded histograms). Whilst (likely) restarted radio-galaxies have been found out to $z\sim3$, the candidate remnant radio-galaxies [with SCI$_0$ $>0.15$] are more-restricted in redshift range. This could be because host-galaxy identification is more difficult for remnant sources than the general population of radio galaxies \citep[e.g. see][and the discussion by Stewart et al., in prep.]{Mahatma2018}, and becomes even more difficult with the fainter core radio-emission at higher redshifts. This is convolved with the difficulty in obtaining redshifts for optically-fainter sources, where the amount of dust and/or distance become limiting factors. Furthermore, for example,  \citet{Godfrey2017} and \citet{Hardcastle2018} suggest that the fraction of remnants is likely to be much lower at higher redshift due to a combination of observational (surface-brightness dimming) and physical (increased
inverse-Compton losses) effects.

By way of concluding Section~\ref{sec:spectral_curvature}, {\color{black} we note that further broadband spectral analysis will be presented in G4Jy Paper V (White et al., in prep.).}

%Other observed-frame properties ?

%Spectroscopy allowing us to de-redshift back to the rest-frame, and look at curvature there. Get Ross Turner and Stas Shabala to help with the interpretation here. [Or should that be a separate paper?]

\subsubsection{Radio luminosities}
\label{sec:luminosities}

In Paper III (figure 9) we presented the radio-luminosity--$z$ distribution of G4Jy sources that have newly-collated redshifts in the updated G4Jy catalogue (Section~\ref{sec:catalogue}). This showed that, as predicted, the lower flux-density threshold used to select the G4Jy Sample \citep{White2020a,White2020b}, compared to 3CRR \citep{Laing1983}, allows both lower radio-luminosities and higher redshifts to be probed. This is important for future studies of the interplay of AGN activity and star formation in the host galaxy, with the large sample size also allowing environmental effects to be taken into consideration.  

\begin{figure*}
\centering
\subfigure[Radio luminosities of G4Jy-NVSS sources at 151\,MHz, by spectral curvature (SCI$_0$)]{
\includegraphics[scale=0.8]{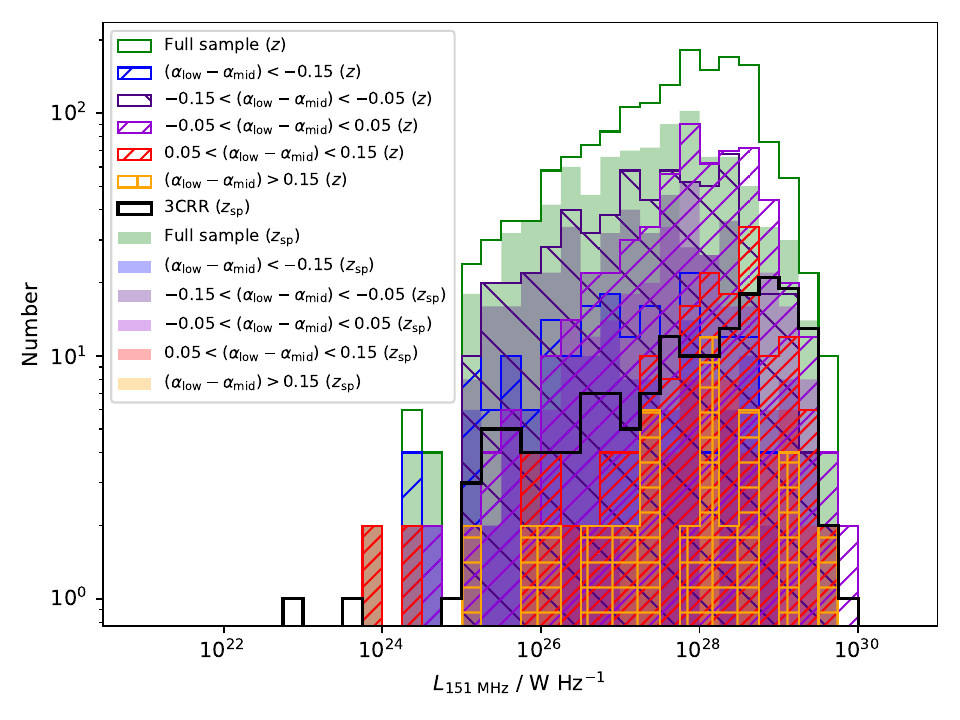} }
\subfigure[Radio luminosities of G4Jy-NVSS sources at 1400\,MHz, by spectral curvature (SCI$_0$)]{
\includegraphics[scale=0.8]{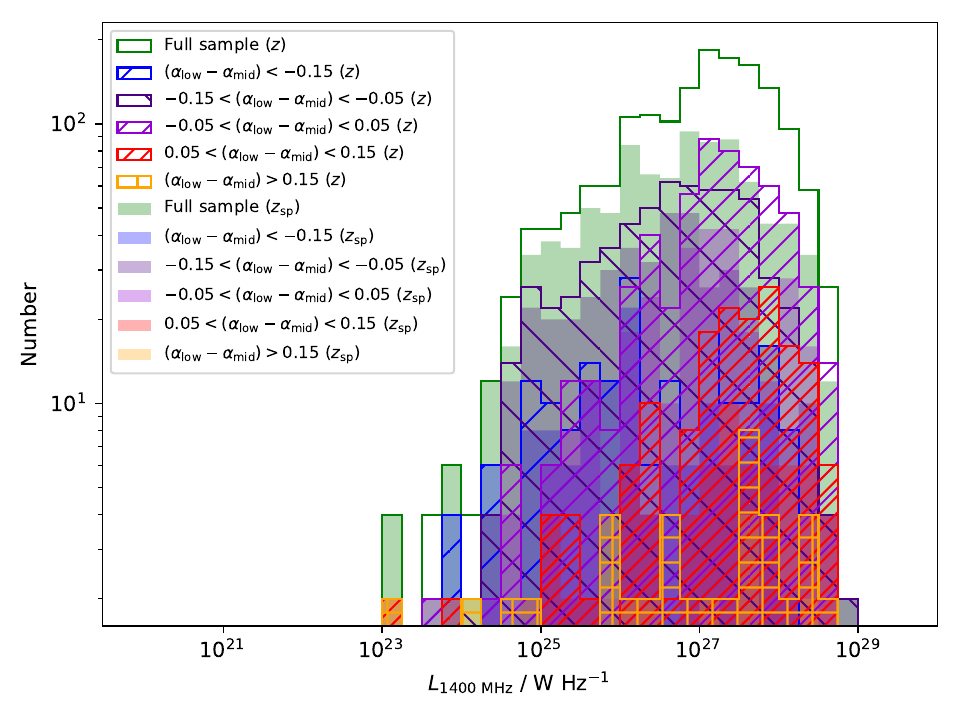} \label{fig:scatterplot}}
\caption{Distributions in the radio luminosities at (a) 151\,MHz, and (b) 1400\,MHz (Section~\ref{sec:luminosities}), colour-coded by the (radio) spectral curvature of the source (Section~\ref{sec:spectral_curvature}). The luminosity distribution for 3CRR sources \citep{Laing1983} is added (to panel a) for comparison (black, unfilled histogram). In addition, `cumulative redshifts' ($z = z_{\mathrm{sp}}$, $z_{\mathrm{ph}}$, $z_{\mathrm{NED}}$) are used for the outlined histograms, whilst the shaded histograms are based on only the $z_{\mathrm{sp}}$ values being used for the luminosity calculation.  }
\label{fig:luminosities}
\end{figure*}

For this work,  we present the radio luminosities of G4Jy-NVSS sources as a function of spectral curvature, SCI$_0$, in Figure~\ref{fig:luminosities}. This is a novel way of investigating the properties of the sources, but their luminosity distributions do not show any discernible trends. For example, we may have expected the (candidate) `restarted' radio-galaxies (in blue) to show higher-than-average radio luminosities, but the actual distribution (Figure~\ref{fig:luminosities}a) is a reminder that the {\it low-}frequency radio emission is tracing {\it older} populations of relativistic electrons. To see whether our expectation holds for 1400-MHz radio-luminosities, which tend to trace more-recent AGN activity, we plot the latter in Figure~\ref{fig:luminosities}b. The only marginal trend that can be seen is the slightly greater clustering of $0.05 < \mathrm{SCI}_0 < 0.15$ sources (in red) towards higher 1400-MHz luminosities. It is possible that this is connected to the relativistic beaming that is seen at higher radio-frequencies, but if that were the case then we would expect to see a similar effect for sources with more negative values of spectral curvature. Remnant radio-galaxies are `exempt' from this expectation because the 1400-MHz luminosity may still be confined to the unbeamed radio-lobes, as opposed to the beamed radio-core (itself depending on the orientation of the radio-jet axis with respect to the line-of-sight).  

Of course, a complicating factor for all radio luminosities is the strength of the magnetic field within which the relativistic electrons (giving rise to the radio emission) are accelerating. We hope that such degeneracy will be addressed through 
X-ray observations of a complete, and statistically-large, radio-source (sub)sample.

\subsubsection{Linear sizes} 
\label{sec:linearsizes}

Again, as demonstrated in Paper III, the newly-acquired redshifts allow us to calculate the linear sizes of G4Jy sources. These are derived from angular sizes based on 45-arcsec imaging (see section~5.2 of Paper III), which may lead to overestimates of the extent of the radio emission. Such overestimation will affect small sources more greatly than larger sources, with projection effects further complicating the angular-size distribution.

\begin{figure}
\centering
\vspace{-0.5cm}
\subfigure[Linear sizes by spectral curvature (SCI$_0$)]{
\includegraphics[scale=0.5]{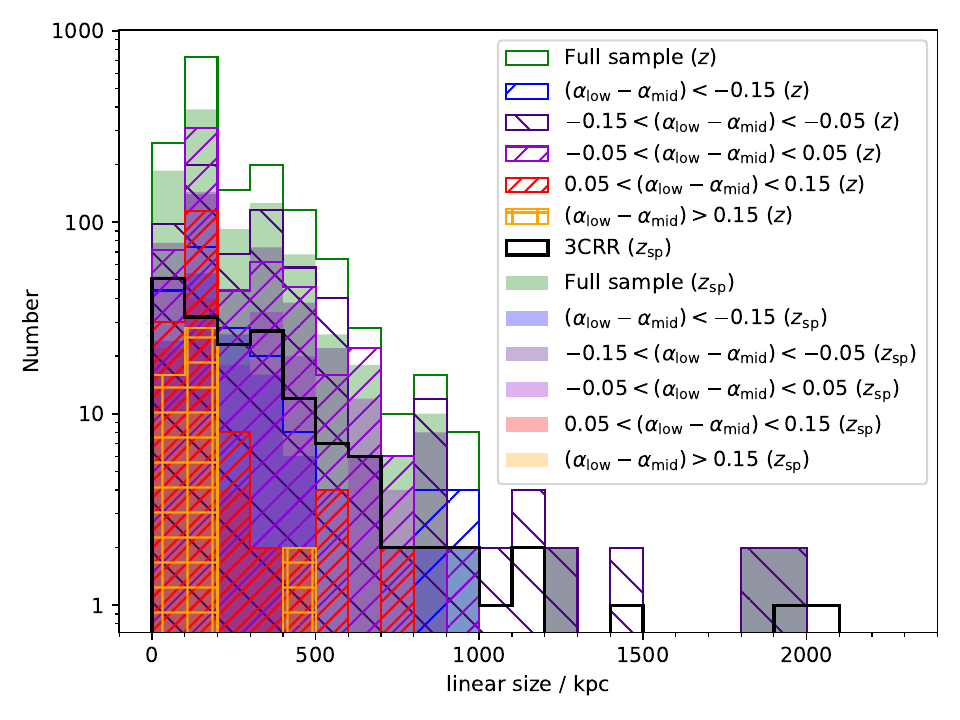} \label{fig:scatterplot}}
\caption{ Distributions in the linear sizes of G4Jy-NVSS sources (Section~\ref{sec:linearsizes}), colour-coded by the spectral curvature index, SCI$_0 = \alpha_{\mathrm{low}} - \alpha_{\mathrm{mid}}$ (Section~\ref{sec:spectral_curvature}). The size distribution for 3CRR sources \citep{Laing1983} is added for comparison (black histogram), with one source (3C\,236) having a linear size that is beyond the plot range (at 4530\,kpc). The outlined histograms represent linear sizes calculated from the `cumulative redshifts' ($z = z_{\mathrm{sp}}$, $z_{\mathrm{ph}}$, $z_{\mathrm{NED}}$), whilst the shaded histograms represent linear sizes calculated from spectroscopic redshifts alone ($z_{\mathrm{sp}}$). }
\label{fig:linearsizes}
\end{figure}
% (NGC\,6251, 3C\,326 = G4Jy\,1282, and 3C\,236) having linear sizes that are beyond the plot range (i.e. 1900\,kpc to 4530\,kpc)

For interest, we present the linear sizes as a function of spectral curvature (Figure~\ref{fig:linearsizes}). As with the corresponding radio-luminosity figure, we are able to discern few trends with respect to SCI$_0$ (Equation~\ref{eq:sci0}). One exception is that the (likely) oldest population (i.e. sources with SCI$_0 > 0.15$) have linear sizes below 600\,kpc. This could be due to a scenario in which these sources happen to be more-confined by their environment (or growing more slowly) than is the case for the general population. Alternatively, it is the result of a selection effect, where more-extended radio emission (possibly from previous episodes of AGN activity) is beyond the sensitivity limit of the NVSS data, and so the linear size is underestimated. (As a reminder, only G4Jy sources within the NVSS footprint are considered, in terms of spectral curvature, because these have an $\alpha_{\mathrm{mid}}$ value calculated between 151 and 1400\,MHz.) 

Consideration of which sources cross the 1-Mpc threshold to be categorised as giant radio galaxies (GRGs) is detailed in Appendix~\ref{sec:GRGs}.

\subsubsection{$P$--$D$ diagrams}
\label{sec:PDdiagrams}

\begin{figure*}
\centering
\includegraphics[scale=1.0]{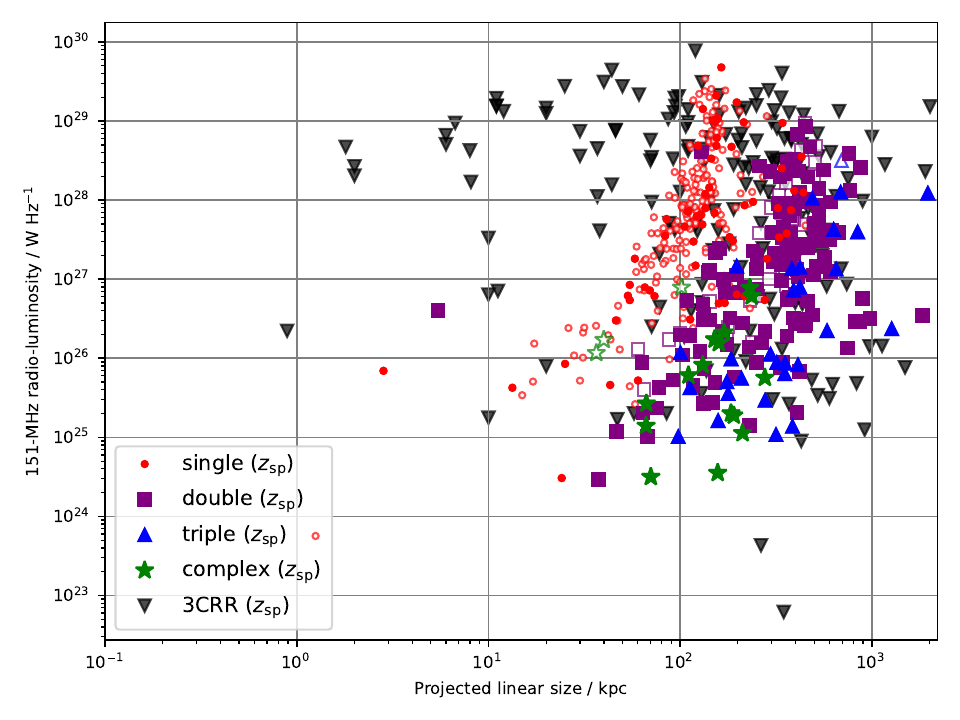}
\caption{ The $P$--$D$ diagram \citep{Baldwin1982} for the G4Jy Sample (Section~\ref{sec:PDdiagrams}), thanks to newly-acquired spectroscopic redshifts for 34 per cent of the sample {\color{black}\citep[][]{White2025a,White2026}}. Datapoints are colour-coded by the radio morphology of the sources, as determined by \citet{White2020a,White2020b} via visual inspection of NVSS/SUMSS imaging (of 45-arcsec resolution). Also plotted, as black triangles, are sources belonging to the 3CRR sample \citep{Laing1983} for comparison (and as a visual aid).  The higher-resolution radio-maps available for 3CRR sources {\color{black}explain the lower linear-sizes that they reach, compared to the G4Jy Sample. Unfilled} symbols represent G4Jy sources for which the linear-size measurement (via 45-arscec imaging) is an upper limit.  }
\label{fig:PDdiagram}
\end{figure*}

\begin{figure*}
\centering
%\vspace{-0.5cm}
\subfigure[$P$--$D$ diagram colour-coded by $z_{\mathrm{sp}}$]{
\includegraphics[scale=0.85]{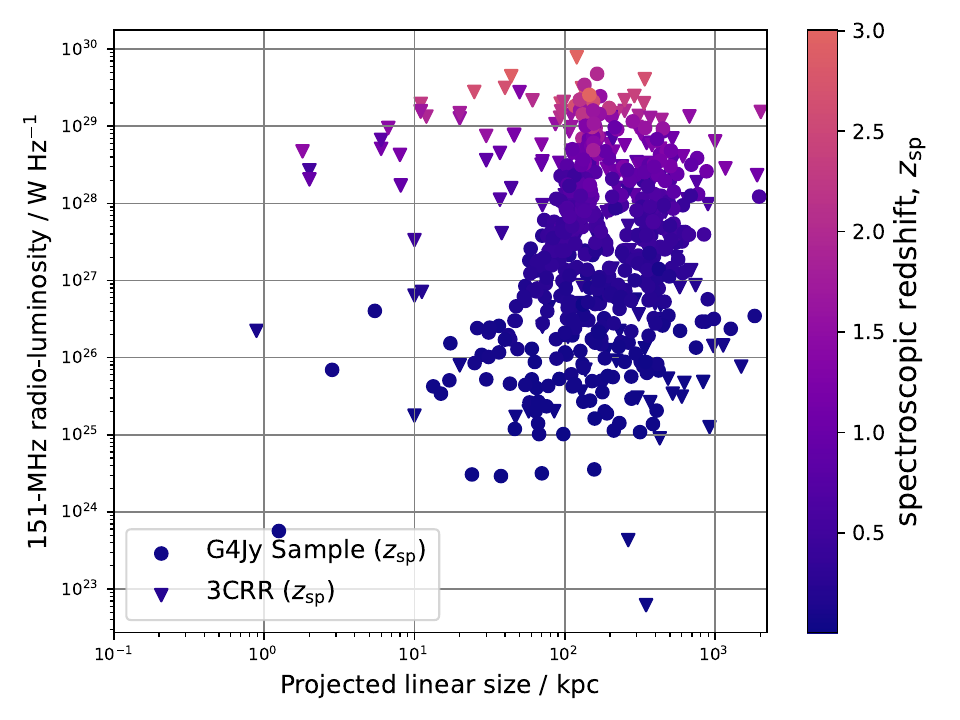} }
\subfigure[$P$--$D$ diagram colour-coded by spectral curvature, SCI$_0$ (with unfilled symbols representing linear-size upper-limits)]{
\includegraphics[scale=0.9]{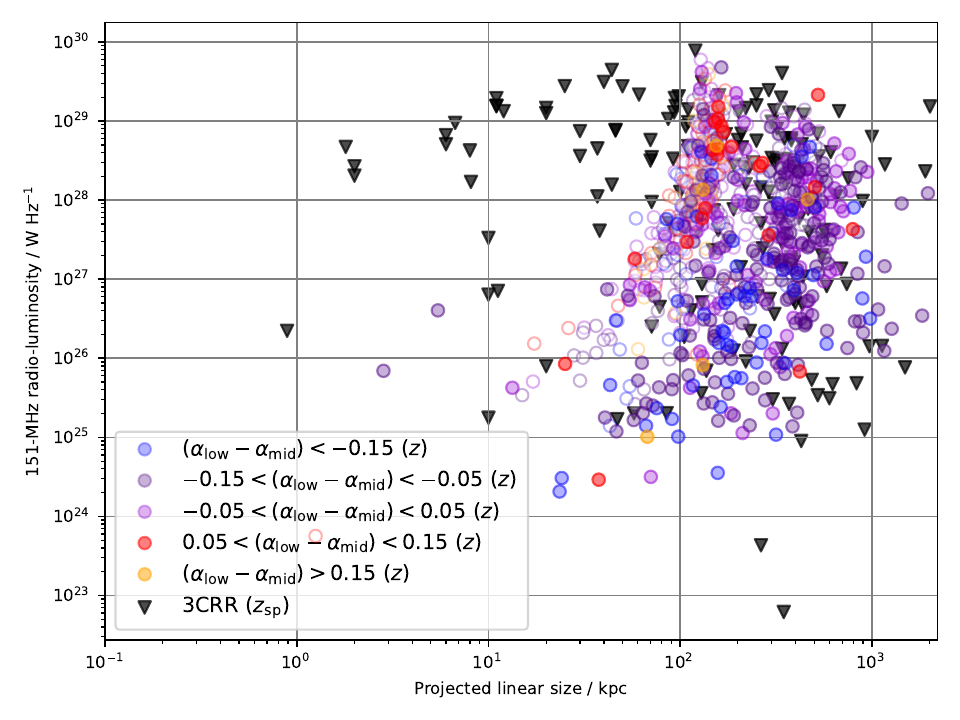} \label{fig:scatterplot}}
\caption{ $P$--$D$ diagrams for G4Jy-NVSS sources (Section~\ref{sec:PDdiagrams}), where the 151-MHz radio luminosity (`radio power', $P$) is plotted against the linear size ($D$). The two panels are colour-coded by (a) spectroscopic redshift, and (b) spectral curvature, SCI$_0 = \alpha_{\mathrm{low}} - \alpha_{\mathrm{mid}}$ (Equation~\ref{eq:sci0}). The distribution for 3CRR sources \citep{Laing1983} is added for comparison (triangles), with 3C\,236 having a linear size that is beyond the plot range, at 4530\,kpc. The number of sources is higher for panel (b), where $z$ refers to the `cumulative redshifts' ($z_{\mathrm{sp}}$, $z_{\mathrm{ph}}$, $z_{\mathrm{NED}}$) that have been compiled for the sample, i.e. avoiding duplication of sources. }
\label{fig:PDdiagrams}
\end{figure*}

In Figures~\ref{fig:PDdiagram} and \ref{fig:PDdiagrams} we present the 151-MHz radio-luminosity against the (projected) linear size. This parameter space is effectively a radio-power--size ($P$--$D$) diagram, as first constructed by \citet{Baldwin1982} to help elucidate the evolutionary tracks of extended radio-galaxies. (See also \citealt{Ryle1967}, and note that we have additionally plotted compact radio-galaxies in our figures, for completeness.) It is believed that radio galaxies traverse this $P$--$D$ plane during their lifetime, expanding and brightening-then-fading as they do so, with starting points towards the left-hand side of the diagram and endpoints towards the right-hand side. The shape of their path is governed by the jet power and the environment into which the radio emission is expanding, which themselves are complicating factors (given the variety of jets and environments that we witness in radio-galaxy populations; e.g. \citealt{Vardoulaki2024}). 

Another complication is estimating how quickly hotspots advance through the intergalactic medium \citep[e.g.][]{Alexander1987,An2012,Stewart2025a}. However, this is only relevant for sources with FR-II morphology, and so simulations have been extended to incorporate the necessary physics for modelling FR-I radio-galaxies in addition \citep[e.g.][]{Turner2018,Turner2023}. This includes the difference in spectral signatures of the lobes of FR-Is and FR-IIs, caused by the different directions in which the shock-accelerated particles flow from the flare point. (That is, `forwards' for FR-I sources and `backwards' for FR-II sources.) We note that the viewing angle of the source will also have an impact on the spectral signature, with radio-jets close to the line-of-sight having their (high-frequency) flux-densities Doppler boosted, resulting in a flatter-than-average radio spectrum. This is less of a concern for the G4Jy Sample, since the low-frequency selection of these sources results in a sample that is unbiased with respect to the orientation of the radio-jet axis \citep{Barthel1989}.

Despite all of these complex factors, the $P$--$D$ diagram offers an informative way of exploring the radio-loud AGN population in further detail (e.g. see the model tracks in figure 8 of \citealt{Hardcastle2019}). The G4Jy Sample has a similar spread in radio luminosities (Figure~\ref{fig:luminosities}) and linear sizes (Figure~\ref{fig:linearsizes}) as the 3CRR sample \citep{Laing1983}, and we now draw attention to the correspondence of different radio morphologies \citep{White2020a,White2020b} to different regions of $P$--$D$ space (Figure~\ref{fig:PDdiagram}). 

Firstly, we note that the `complex' label was assigned by \citet{White2020a,White2020b} to sources with unusual morphologies (usually in cluster environments) but also to head-tail radio-galaxies and narrow-angle tailed radio-galaxies (whose morphologies are believed to have formed from the relative motion of the source through the cluster medium -- see Paper II for examples). These sources tend to have lower radio-luminosities than the `doubles' in our sample, and they occupy the lower left-hand region of $P$--$D$ space. This suggests that the stability and density of the surrounding medium may restrict their distribution to small linear sizes. However, we know that `complex' morphology can be maintained on large scales \citep[e.g.][]{Jaffe1973}, meaning that the small extents of `complex' G4Jy sources is more-likely a selection effect (imposed by the surface-brightness limit of the radio imaging that is involved).  

Meanwhile, the distribution of `complex' sources appears to be in agreement with the work of \citet{Chilufya2025}, who also study the $P$--$D$ diagram as a function of radio morphology. They use the 2.5-arcsec resolution provided by VLASS to help with visual inspection of extended radio galaxies ($>$60\,arcsec) detected in LoTSS DR2 \citep{Shimwell2022}. Their wide-angle tail, narrow-angle tail, head-tail, and relaxed-double radio-sources all occupy the same region of $P$--$D$ space as our `complex' sources, as do their radio galaxies showing typical FR-I and FR-II morphologies. Interestingly, this region is dominated by LERGs \citep[see figure 7 of][]{Chilufya2025}, although this is likely influenced by selection effects -- notably their restriction to $z < 0.57$ -- since LERGs tend to dominate the radio-galaxy population at low redshifts \citep{Best2012}. 

Quite striking is the distribution in $P$--$D$ space of the sources with `single' morphology (Figure~\ref{fig:PDdiagram}), which (unfortunately) are not considered for the sample of \citet{Chilufya2025}. These can generally be divided into two groups: firstly, the compact/point-like radio-sources that follow a trunk-like distribution. This can be explained by: (i) noting that the linear sizes for many of these sources are upper limits, given the spatial resolution of the radio data. (It is worthwhile comparing Figure~\ref{fig:PDdiagram} with figure 7 of \citealt{White2025a}.) (ii), these may be sources whose radio jets have become `frustrated' by the surrounding medium, slowing their maturation into (more-familiar) extended radio-galaxies. This is particularly relevant for short-lived compact symmetric objects (CSOs), as discussed by \citet{An2012}. Secondly, there are `single' sources that are distributed more sparsely across a range of linear sizes, and even passing the commonly-used 700-kpc threshold \citep[e.g. see][]{2005AJ....130..896S} for being classed as a GRG. These are sources that exhibit typical `core-brightened', FR-I-like radio morphology, {\it and} FR-II sources that are not spatially resolved in NVSS/SUMSS. This is confirmed in higher-resolution radio images from MeerKAT \citep{Sejake2023} and VLASS (Dong et al., in prep.).

Naturally, `edge-brightened' FR-II sources fall most easily into the `double' category, and they span the largest range in $P$--$D$ space (Figure~\ref{fig:PDdiagram}, see also figure 6 of \citealt{Chilufya2025}). This suggests that a combination of jet power and environmental conditions have promoted their growth, with lifespans that are long enough that these sources may demonstrate renewed activity (Stewart et al., in prep.). The same can be said for sources with `triple' morphology, although there is a slight clustering of these sources in the same region as the `complex' sources. This could again be due to selection effects: for their `triple' morphology to be distinguished they most likely are at lower redshift, where Malmquist bias means that they are also likely to have lower radio luminosities than the general population. This should be investigated further with higher-resolution radio images for the full sample. This selection effect is also evident in Figure~\ref{fig:PDdiagrams}a, where we see the expected trend between luminosity and redshift. % This is because, for flux-limited samples, there is a tendency for higher luminosities to be probed via higher redshifts.}

In addition, interpretation of the distribution of radio galaxies in the $P$--$D$ diagram is complicated by the {\color{black}infrequently-studied} strength of the magnetic fields within which the relativistic electrons are accelerating \citep[e.g.][]{Miley1980,Jamrozy2004,Ineson2017}. This will influence both the intrinsic jet-power and the path of the radio jets through the surrounding medium \citep{Hardcastle2014}. To the best of our knowledge, how magnetic fields may contribute to slowing the radio jets to small spatial scales is yet to be investigated for a complete AGN sample. However, this could help to explain the distribution of (candidate) ageing radio-sources (SCI$_0 > 0.05$) in Figure~\ref{fig:PDdiagrams}b (which is the first time that the $P$--$D$ diagram has been studied as a function of spectral curvature). Whilst some candidate remnant radio-galaxies can be seen towards the right-hand side of the plane, where evolutionary-track `endpoints' lie, there is a predominance of these sources below $\sim$200\,kpc. This could represent a typical spatial scale beyond which these sources are less able to penetrate the circumgalactic medium and transition into more-extended radio-galaxies \citep[e.g.][Stewart et al., in prep.]{Turner2023}. When we have better radio-morphology information, it would be interesting to see whether there is a correspondence between the FR-I/FR-II morphologies and their distribution in [$P$--$D$](SCI$_{0}$) space. However, the caveat mentioned in Section~\ref{sec:spectral_curvature} remains: the small linear sizes of SCI$_0 > 0.05$ radio galaxies may be due to the young age of the source, where not enough time has passed for the radio jets to have travelled larger distances. 

% Referee said to remove this: footnote{As noted in the review by \citet{Tumlinson2017}, \citet{Prochaska2017} measure the cool, circumgalactic medium out to 160\,kpc for $z \sim 0.2$ quasars, using multiple sightlines with the {\it Hubble Space Telescope}.}

\setkeys{Gin}{draft}

The fates of compact symmetric objects (CSOs) could also hold the key to understanding the predominance of (candidate) ageing radio-sources with linear sizes below $\sim$200\,kpc. As explained by \citet{An2012}, CSOs are observed as either the young stage of all radio sources (some of which expand into FR-I/FR-II morphologies), or are the endpoints of sources whose radio-jets were unable to propagate beyond the boundary between the interstellar medium (ISM) and the intergalactic medium (IGM). The typical length scales for such sources are on the order of 0.001--1\,kpc \citep[][see their figure 1]{An2012}, where the ``short-lived central source activity results in flow instabilities before the ISM--IGM transition point, and the eventual dissipation of the lobe structure''. 

Furthermore, by considering the unfilled symbols in Figure~\ref{fig:PDdiagrams}b, we can interpret the slight positive correlation between luminosity and size (for this `limb' of SCI$_0 > 0.05$ sources) as upper limits imposed by the spatial resolution of the NVSS imaging. By obtaining more-accurate angular-sizes (and thereby linear sizes) for the full sample, the region where 3CRR sources are visible in the top left-hand corner of the $P$--$D$ diagram will also be populated by G4Jy sources. This is relevant for FR-I sources in general, whose angular sizes may have been severely mis-estimated through the relatively poor brightness-sensitivity of NVSS compared to, e.g., MeerKAT. Further work (beyond the scope of this paper) would also help to determine the significance of projection effects on the linear-size distribution of the ageing radio-sources.  

We also note the unusual prevalence of large radio-galaxies showing spectra that flatten at mid-frequencies (SCI$_0 < -0.05$). This is indicative of renewed AGN activity, with these sources possibly having expanded for so long that their duty cycle has been completed (i.e. these AGN are returning to their `on' state; e.g. see figure 1 of \citealt{Stewart2025a}). They may also have a predisposition for being located `in the field' (allowing for less-constrained expansion than in e.g. clusters), with more-massive-than-average gas reservoirs that enable the AGN to be reignited more-easily. Moreover, spectral flattening at higher radio-frequencies is seen in simulations by \citet{YatesJones2022} for large radio-galaxies at high redshifts (see their figure 14). In this case, such spectral shape arises as a result of the low-frequency  component, associated with the lobes, appearing faint (largely due to inverse-Compton losses) and moving to even lower radio-frequencies ({\it below} 151\,MHz). Meanwhile, the mid-frequency emission associated with the core, jets, and hotspots remains prominent. This will be explored further in future work.

As suggested by \citet{White2025a}, X-ray and polarimetric follow-up observations could allow the degeneracy of some of the complicating factors (mentioned above) to be broken, enabling easier interpretation of the $P$--$D$ diagram. Of course, this would also need to be accompanied by theoretical modelling, in order to understand the physics that underpins the evolution of different types of radio galaxy. Furthermore, after obtaining stellar masses for the G4Jy Sample, we will be able to test whether the steep radio spectral-curvature of more-compact sources corresponds to the `cross-over' \citep{Wilde2023} between the circumgalactic medium and the intergalactic medium.

\subsection{Mid-infrared properties}

\subsubsection{{\it WISE} colour-colour space}
\label{sec:WISE_diagrams}

\begin{figure}
\centering
\includegraphics[scale=0.52]{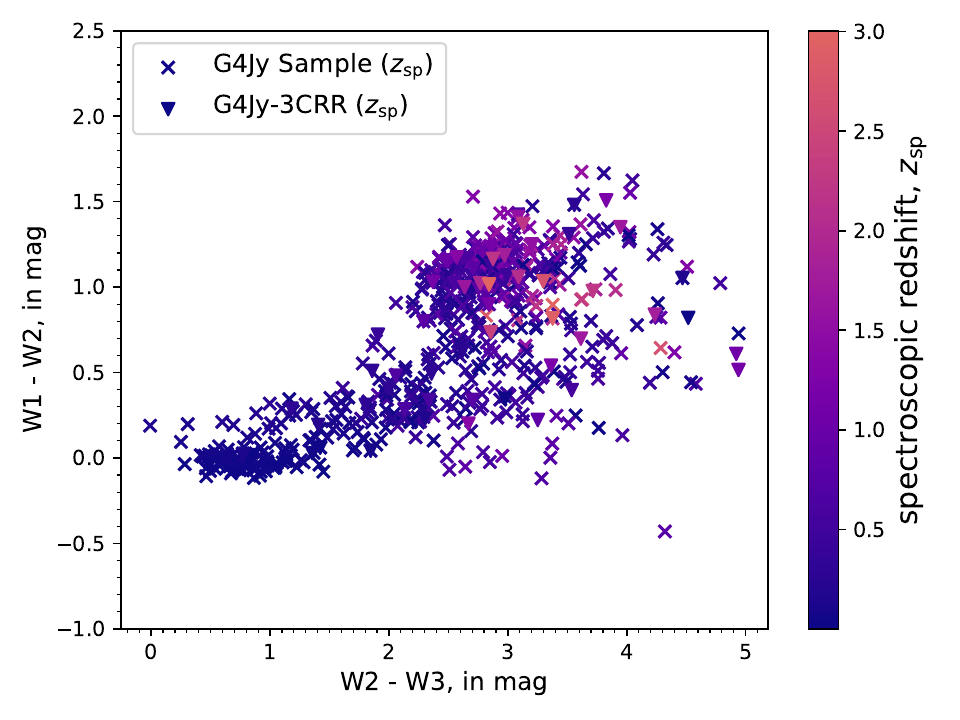}
\caption{The {\it WISE} colour-colour diagram, with markers colour-coded by the spectroscopic redshift. `G4Jy--3CRR' refers to the 67 3CRR sources \citep{Laing1983} that overlap with the sample, as described by \citet{White2020b}.}
\label{fig:WISE_by_specz}
\end{figure}

In this section we focus on {\it WISE} colour-colour space (e.g. Figure~\ref{fig:WISE_by_specz}), which is constructed by plotting W1$-$W2 against W2$-$W3 (see Section~\ref{sec:MIR_data}; \citealt{Wright2010,Jarrett2017}). It has been demonstrated by \citet{Stern2012} that W1$-$W2 > 0.8\,mag can be used as a diagnostic for identifying sources with a warm/hot dusty torus, which are typically found in radiatively-efficient AGN such as quasars. Meanwhile, increasing (i.e. redder) values of W2$-$W3 show a correlation with the rate of star formation in the host galaxy \citep{Jarrett2017}, with starbursts typically found in the bottom right-hand corner of this parameter space.

\setkeys{Gin}{draft=false}

These delineations have led many authors to use mid-infrared colours to differentiate AGN from star-forming galaxies, but the distribution of the G4Jy Sample (for example; see also {\color{black}figure 3 of \citealt{Gurkan2014} and} figure 14 by \citealt{Jarrett2017}) across a wide range of different colour combinations shows that relying on such a method is unwise\footnote{Of course, the criteria used to select AGN are usually a balance between reliability (how do we minimise contamination from other types of source?) and completeness (have all of the AGN been selected?), and so the path taken depends on the science goals. In their figure 6, \citet{Stern2012} demonstrate the tradeoff between reliability and completeness for the W1$-$W2 criterion.}. This is because the vast majority of the G4Jy Sample are {\it known} to be radio-loud AGN, with notable exceptions being two very nearby star-forming galaxies (close enough to have their 151-MHz flux-density passing the 4\,Jy threshold) and a lensed galaxy (where the magnification has boosted both the AGN and star-formation components) -- see \citet{White2020a} for further details. Nonetheless, we note that the AGN signature in our sample is based upon low-frequency radio-emission, which may have been generated on the order of $\sim$100\,Myr ago (\citealt{Alexander1987,Stewart2025a}) if the AGN has since `switched off'. Therefore, the mid-infrared emission may provide insight into {\it more-recent} levels of AGN and star-formation activity, with the caveat that radiatively-inefficient AGN (/low-excitation radio-galaxies; LERGs; \citealt{Best2012}) still populate the region where W1$-$W2 < 0.8\,mag. 

Considering the full G4Jy Sample where there is coverage in W1 and W2, the percentage of sources with W1$-$W2 > 0.8\,mag is ($728/1693=$) 43 per cent, illustrating the need for colours beyond this range to be considered for AGN-selection completeness. (Only one fewer source does not also have coverage in the W3 band, which is of direct relevance for {\it WISE} colour-colour space.) Their median $r$-band magnitude is $r=$\,19.4\,mag, whilst the median $r$-band magnitude for sources with W1$-$W2 < 0.8\,mag is $r=$\,19.9\,mag. This is in good (qualitative) agreement with our initial spectroscopic follow-up \citep[][Sejake et al., in prep.]{GarciaPerez2024,White2025a}, which aims to achieve spectroscopic completeness to $r\sim20$\,mag. 

In Figure~\ref{fig:WISE_by_specz} we present the subset of the G4Jy Sample that has spectroscopic-redshift coverage. As expected, the highest redshifts are for the quasars \citep{Wright2010}, which are in the region defined by the W1$-$W2 > 0.8\,mag criterion of \citet{Stern2012}. The robustness of this classification will be tested in future work, based on visual inspection of optical spectroscopy, and (in the meantime) we direct the interested reader to figure 7 of Sejake et al. (in prep.). 

%Although the `spiral' region has the {\color{blue}second-smallest occupancy (with 22} per cent of the subset), it is of particular interest because we tend to assume that radio-loud AGN are harboured by massive ellipticals (albeit for good reasons; \citealt{Best2012,Sabater2019}). We also note from Figure~\ref{fig:WISE_by_specz} the tendency of ellipticals to be at low redshift, but caution that this is an observational effect. As shown by Sejake et al. (in prep.), these sources have optical spectra that are absorption-line dominated, and it is more difficult to achieve the required signal-to-noise for distinguishing these features (against the continuum) for sources at higher redshift. 

% This suggests that we could be witnessing normal star-forming galaxies in the process of being quenched (REF), with the relative rarity of such systems indicating the rapid timescales over which this quenching takes place (REF). [Am I overinterpreting??] YES :P

\begin{figure}
\centering
%\vspace{-0.5cm}
\subfigure[{\it WISE} colour-colour space, by radio morphology]{
\includegraphics[scale=0.52]{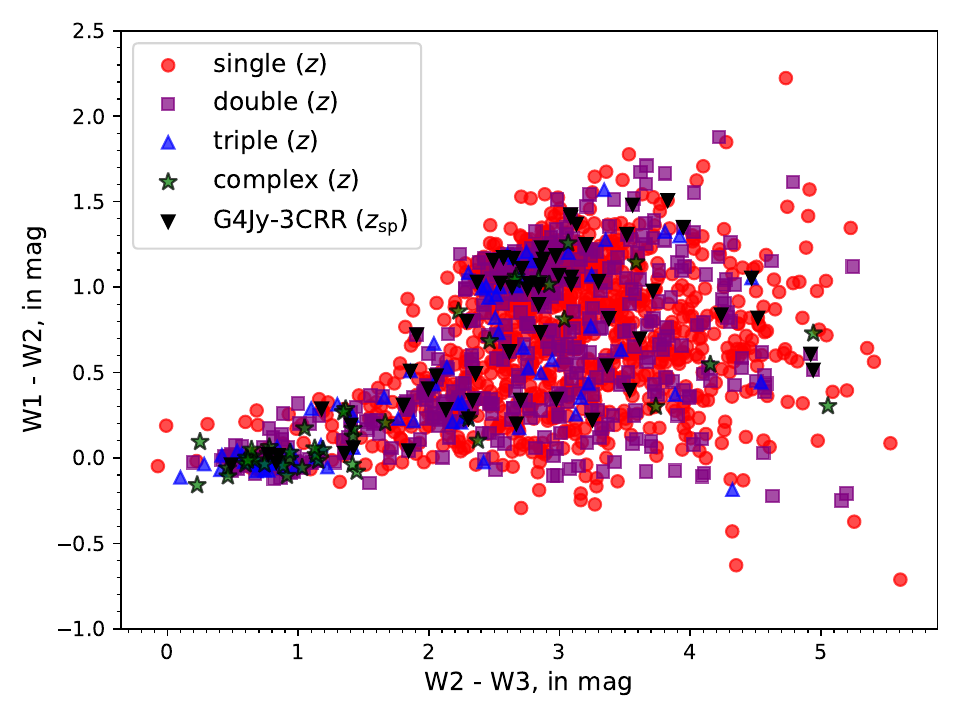} }
\subfigure[{\it WISE} colour-colour space, by spectral curvature (SCI$_{0}$)]{
\includegraphics[scale=0.52]{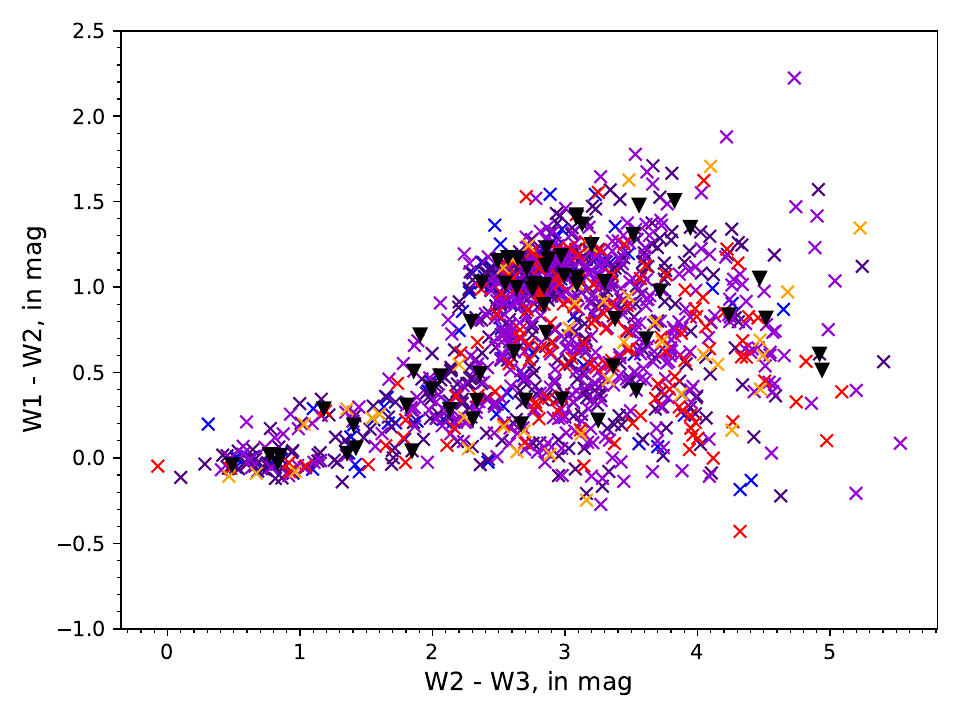} \label{fig:scatterplot}}
\subfigure[{\it WISE} colour-colour space, by optical morphology]{
\includegraphics[scale=0.52]{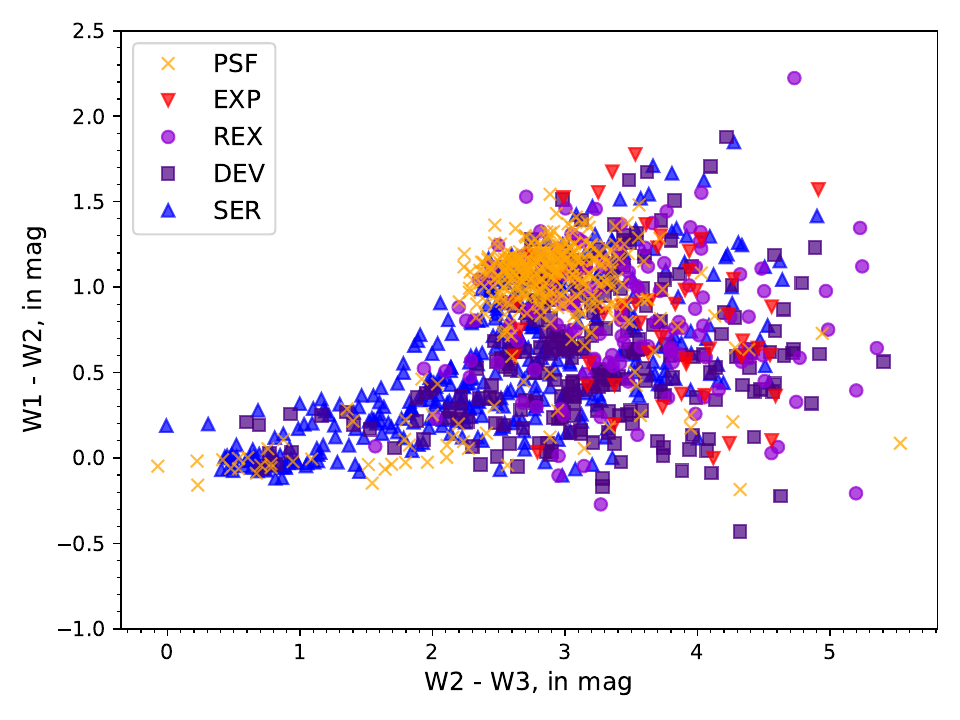} }
\caption{ {\it WISE} colour-colour space (Section~\ref{sec:WISE_diagrams}), with markers representing (a) different NVSS/SUMSS radio morphologies \citep{White2020a,White2020b}, (b) different degrees of spectral curvature (Equation~\ref{eq:sci0}), and (c) the optical morphology of the source (classified via LSDR10; see Table~\ref{tab:sersic_index}). The legend for panel (b) is not shown for size reasons, but the colour-coding is the same as that for Figure~\ref{fig:PDdiagrams}b. }
\label{fig:colourcoded_WISE}
\end{figure}

\setkeys{Gin}{draft}

Next we consider the full G4Jy Sample, in Figure~\ref{fig:colourcoded_WISE}a, where the datapoints are colour-coded by radio morphology. The latter is based on the 45-arcsec imaging from NVSS/SUMSS, and 67 G4Jy-3CRR sources \citep{White2020b} are added for reference. Now, without the criterion that the G4Jy source must have a spectroscopic redshift, we see better population of {\it WISE} colour-colour space. This is largely because optical spectroscopy is biased towards unobscured (Type-1) AGN, whereas the mid-infrared detects both unobscured and dust-obscured AGN very effectively \citep[e.g.][]{Stern2012}. It could be argued that the `single' radio morphology may be concealing additional star-forming galaxies in the sample, but sources with `double' and `triple' morphology are clearly AGN, and found throughout this mid-infrared parameter space. 

%Revisiting the `spirals' region, we observe that the occupancy fraction has increased to {\color{blue}28} per cent. This again suggests that in Figure~\ref{fig:WISE_by_specz} we were witnessing a selection effect, whereby the dust associated with significant star-formation in spiral galaxies \citep[e.g.][]{daCunha2010,Triani2021} means that the optical magnitude will be fainter, making it more difficult to acquire an (optical) spectroscopic redshift. 
%However, common to both figures is the clustering of datapoints in the `QSOs/Seyferts' region and the `ellipticals' region. 
%We will discuss the former region below, and note that G4Jy sources with `complex' radio morphology lie in the latter region. A reminder that this morphology label includes head-tail and narrow-angle-tail radio-galaxies [see \citet{White2020a} for examples], which tend to be found in cluster environments (as do elliptical galaxies; see, e.g., the reviews by \citealt[][]{Cappellari2016,Cappellari2025}, and references therein).

In Figure~\ref{fig:colourcoded_WISE}b we present the same colour-colour space, but {\color{black}for G4Jy-NVSS sources only (accounting for 77 per cent of the G4Jy Sample), and} with markers colour-coded by the degree of spectral curvature (Equation~\ref{eq:sci0}). There appears to be no correlation between the spectral behaviour of the G4Jy sources and their {\it WISE} colours, which suggests a disconnect between the duty cycle of the AGN and the host galaxy. In other words, radio-loud AGN exhibit a wide range of lifespans, independent of their host-galaxy type. {\color{black}Another reason for the number of sources decreasing with respect to Figure~\ref{fig:colourcoded_WISE}a is that, whilst morphology information is available for the full sample, the calculation of SCI$_{0}$ depends on the 85-per-cent availability of a (power-law-fitted) G4Jy\_alpha value. This restriction means that Figure~\ref{fig:colourcoded_WISE}b will be missing G4Jy sources that exhibit even greater degrees of spectral curvature, and those sources will be studied further in G4Jy Paper V (White et al., in prep.). }

We are also interested in the correspondence between the optical morphology (Table~\ref{tab:sersic_index}) and regions of {\it WISE} colour-colour space -- see Figure~\ref{fig:colourcoded_WISE}c. This is most obvious for quasars, which are fitted by PSF light-profiles in the optical, and occupy a well-defined region in terms of mid-infrared colours. Again, the number of sources plotted has decreased {\color{black}(with respect to Figure~\ref{fig:colourcoded_WISE}a)}, this time due to the coverage provided by LSDR10 \citep{Dey2019}, which provides the optical-morphology classifications. We conclude with a note that the three orange (`PSF') crosses close to zero-zero in colour-colour space may actually be misidentified stars that have crept into our host-galaxy identification. These sources are G4Jy~21, G4Jy~636, and G4Jy~1843, and for the latter two, the G4Jy overlays suggest that the host is correct. (We propose MeerKAT follow-up to obtain a higher-resolution radio image for G4Jy 21.)

%The exponential (`EXP') light-profiles are intended for spiral galaxies, but such associations are not limited to the `spirals' region. Similarly, elliptical galaxies are expected to be fitted by S{\' e}rsic-like light-profiles (`DEV', `SER'), and indeed we see those markers populating (but not limited to) that region of {\it WISE} colour-colour space. 

\subsubsection{W1-magnitude distributions}
\label{sec:W1histograms}

\begin{figure*}
\centering
%\vspace{-0.5cm}
\subfigure[Comparing the G4Jy Sample with other samples]{
\includegraphics[scale=0.52]{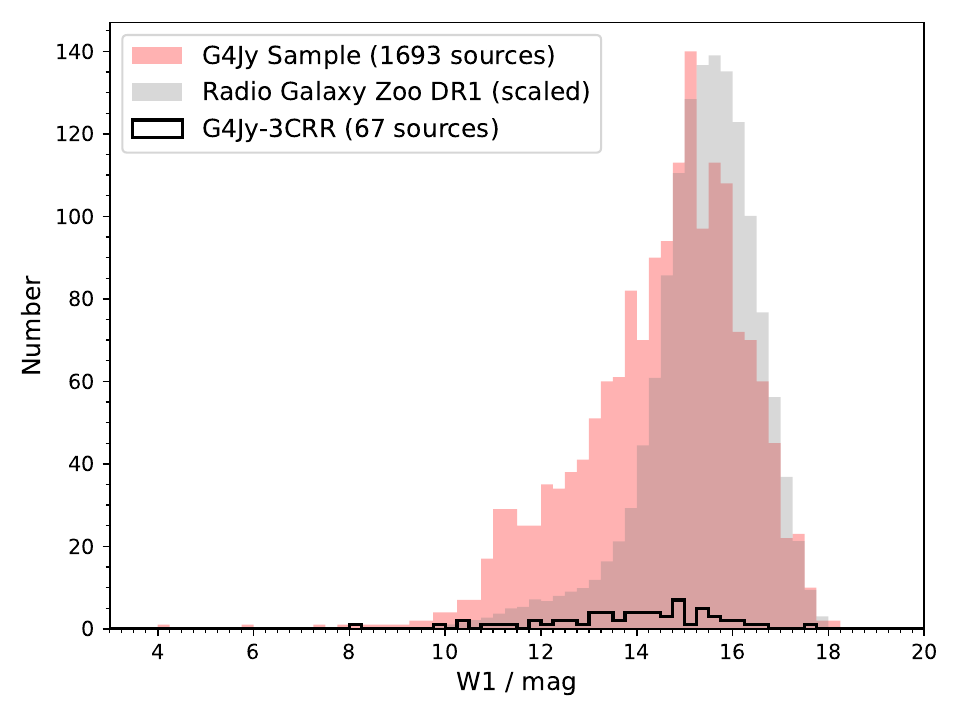} }
\subfigure[The W1-mag distribution colour-coded by morphology]{
\includegraphics[scale=0.52]{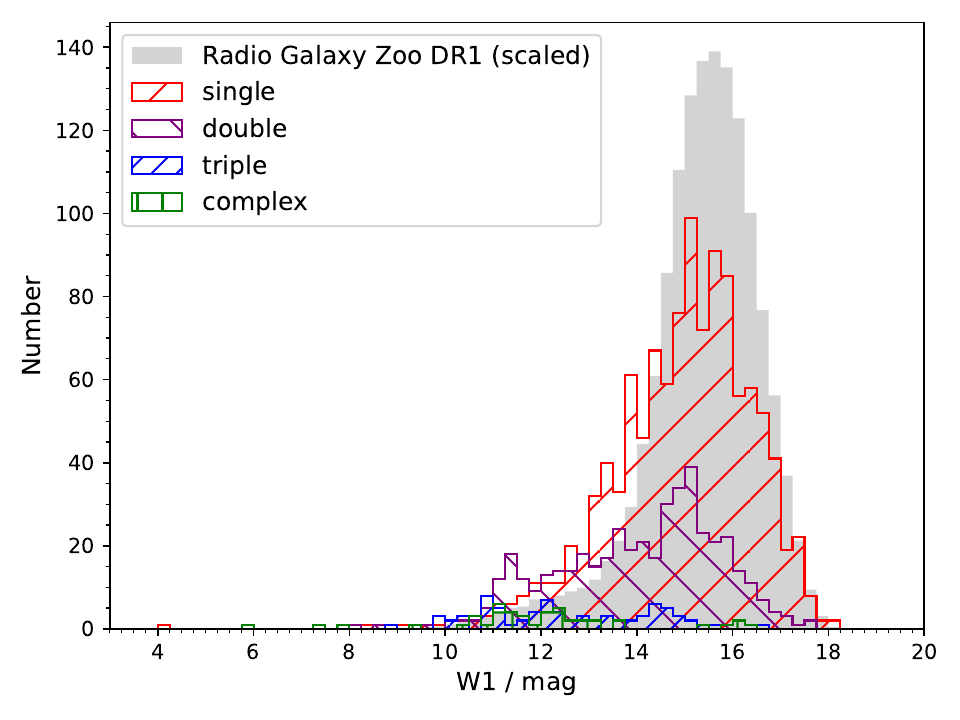} }
\subfigure[The W1-mag distribution colour-coded by redshift bin]{
\includegraphics[scale=0.52]{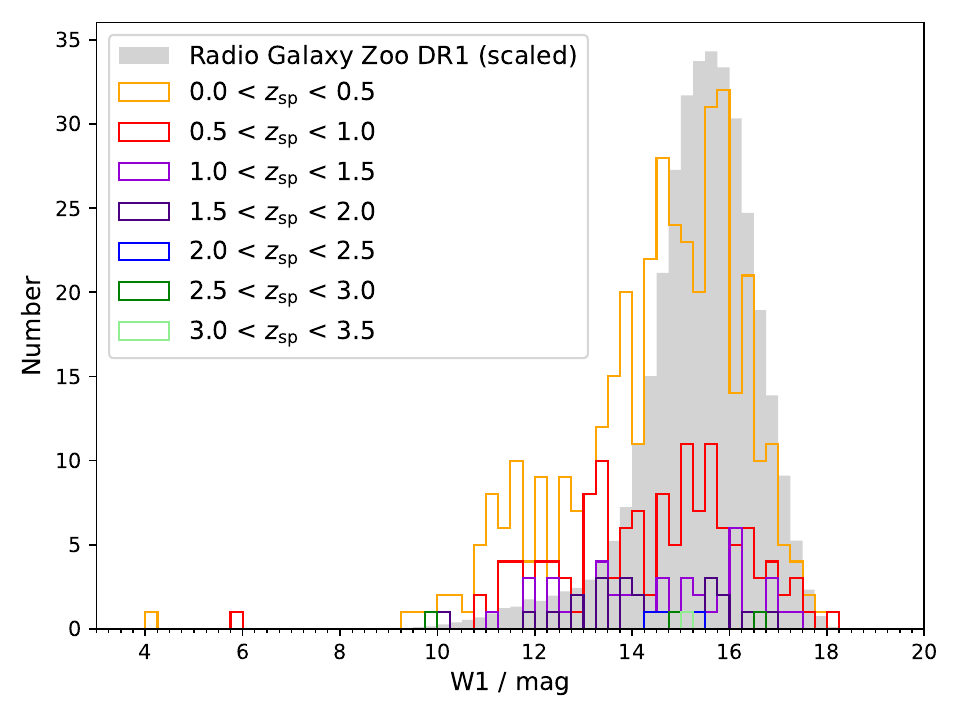} }
\subfigure[Two-Gaussian fitting to the  distribution for $z < 0.5$ sources]{
\includegraphics[scale=0.52]{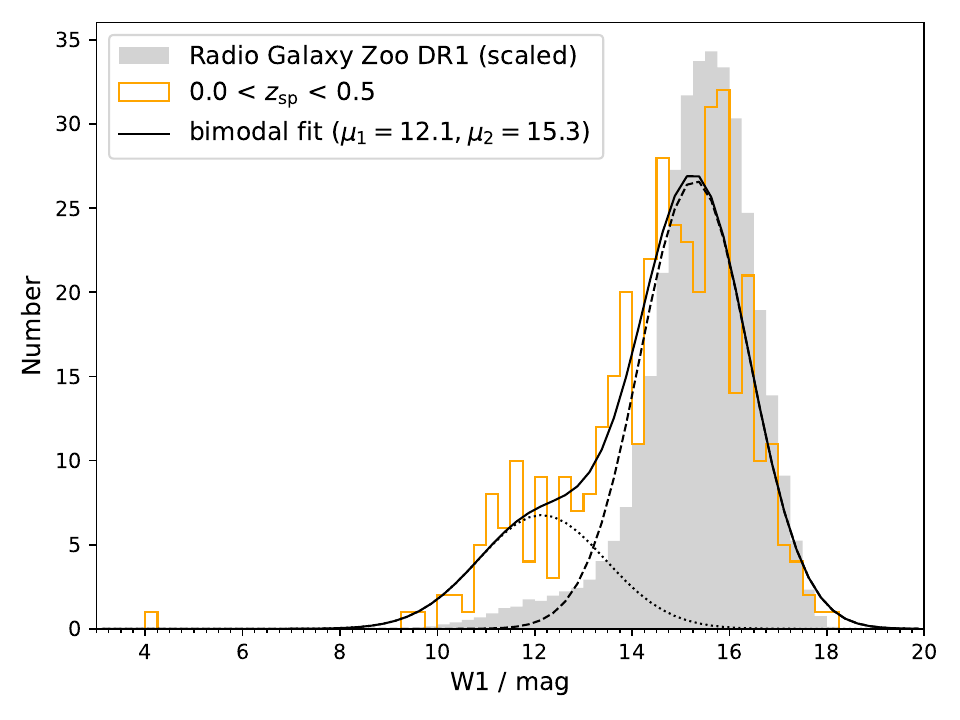} }
\subfigure[The W1-mag distribution for sources hosted by clusters]{
\includegraphics[scale=0.52]{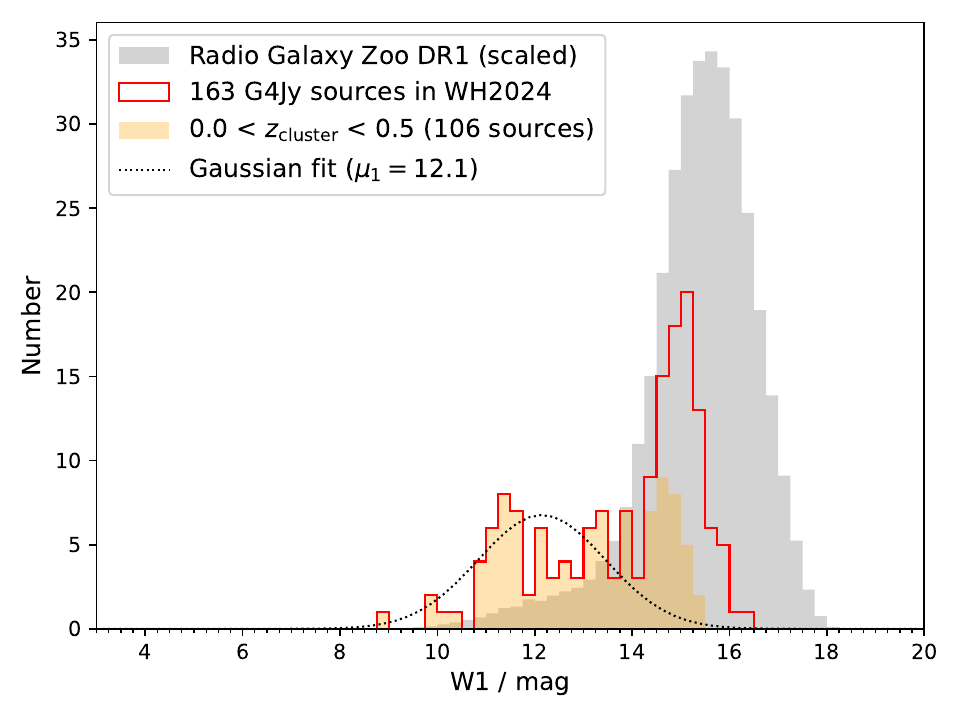} }
\caption{ AllWISE-W1 magnitude distributions for the G4Jy Sample (Section~\ref{sec:W1histograms}). Spectroscopic redshifts are used for panels (c) to (d), to help investigate the origin of the `excess' of brighter W1 magnitudes seen below W1$\sim$14\,mag, with respect to the Radio Galaxy Zoo DR1 \citep{Wong2025} FIRST-based \citep{Becker1995,White1997} catalogue. The two Gaussians that are fitted in panel (d) have mean and standard deviations of $\mu_{1}=12.1$, $\mu_{2}=15.3$ and $\sigma_{1}=1.31$, $\sigma_{2}=1.12$, respectively. Just the first of these Gaussians is retained for {\color{black}panel (e), for clarity, whilst `WH2024' refers to the cluster catalogue  of \citet{Wen2024}. } }
\label{fig:W1histograms}
\end{figure*}

For further interest, we present the W1-magnitude distribution for the G4Jy Sample in Figure~\ref{fig:W1histograms}, with the 67 3CRR sources that overlap with the sample showing a similar range in W1 values. We also show the W1 magnitudes for 55,731 radio galaxies in the (FIRST-based catalogue of the) first data release of Radio Galaxy Zoo \citep{Wong2025}, whose number counts have been downscaled to ease comparison of the overall shape in the distribution. Intriguingly, in Figure~\ref{fig:W1histograms}a, we see an `excess' in the G4Jy Sample (for W1 magnitudes below $\sim$14\,mag) that warrants further investigation.

\setkeys{Gin}{draft=false}

We first see whether this excess has a connection to the radio morphology of the source (Figure~\ref{fig:W1histograms}b). Whilst each of the constituent histograms (for `single', `double', `triple', and `complex' morphologies) show hints of an excess at bright W1 magnitudes, it remains difficult to interpret why this may be the case. A more-informative approach is to split the sample into spectroscopic-redshift bins (Figure~\ref{fig:W1histograms}c), as doing so shows an interesting correspondence for the $0.0< z_{\mathrm{sp}} <0.5$ bin. In fact, as illustrated in Figure~\ref{fig:W1histograms}d, the W1-magnitude distribution for sources at $0.0< z_{\mathrm{sp}} <0.5$ can be modelled by two Gaussian distribution functions: one with a mean of $\mu_{1}=12.1$, corresponding to the `excess', and the other with a mean of $\mu_{1}=15.3$, corresponding to the majority of the W1 distribution.  

Next we {\color{black}consider what could be contributing towards the brighter-magnitude Gaussian ($\mu_{1}=12.1$) that describes the observed excess. The} brighter magnitudes could be the result of these sources being part of merging systems or are in particularly-dense environments such as clusters. We can test the latter by cross-matching the G4Jy Sample with the optical-based cluster catalogue produced by \citet{Wen2024}, using a match radius of 1\,arcsec. This results in 163 cross-matched G4Jy sources (red, unfilled histogram in Figure~\ref{fig:W1histograms}e), 106 of which are at $0.0< z_{\mathrm{cluster}} <0.5$ (orange, filled histogram in Figure~\ref{fig:W1histograms}e). Judging by the distribution of the latter, we have good reason to believe that the excess in bright W1 magnitudes is the result of low-redshift ($z<0.5$) G4Jy sources being harboured by clusters. The reason that this may not be seen in the W1 distribution for FIRST galaxies in Radio Galaxy Zoo is that the FIRST survey \citep{Becker1995,White1997} is not as sensitive towards diffuse emission as GLEAM, and so the hosts of diffuse radio-galaxies (like those found in cluster environments) do not feature.

\subsubsection{Infrared-faint radio-sources}
\label{sec:IFRSs}

\begin{figure}
\centering
\includegraphics[scale=0.52]{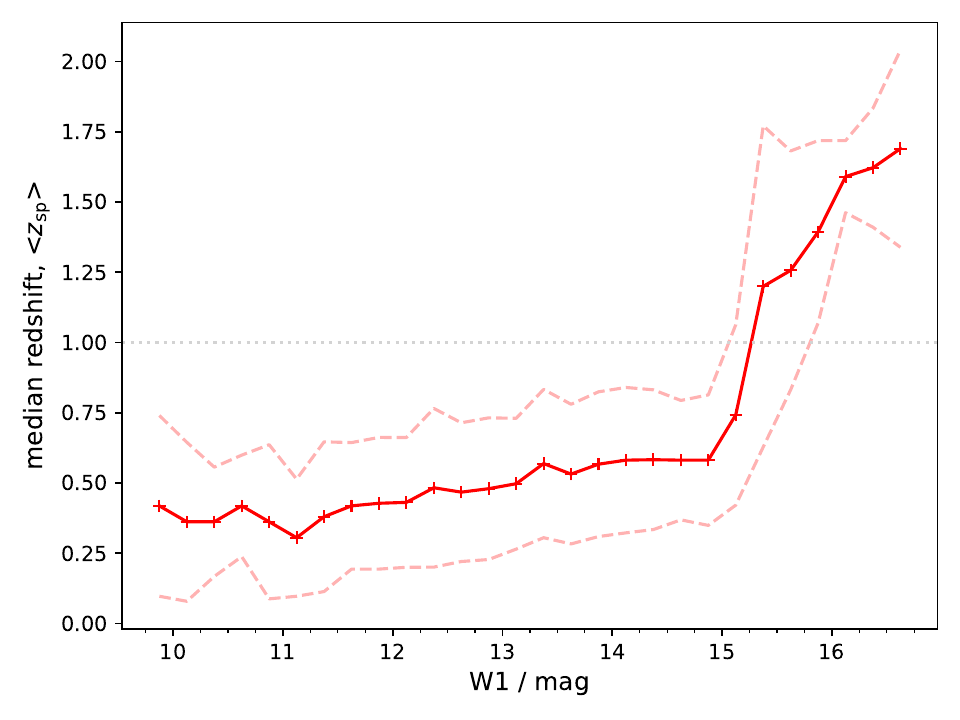}
\caption{The red, solid line represents the median spectroscopic redshift per bin in AllWISE-W1 magnitude (Figure~\ref{fig:W1histograms}), while the pale-red, dashed lines signify its median absolute deviation. This figure indicates that we can expect to find high redshifts ($z > 1$) for sources that are fainter than W1 $\sim$ 15.25\,mag (Vega), as expanded upon in Section~\ref{sec:IFRSs}.}
\label{fig:W1_by_medianspecz}
\end{figure}

Now we briefly consider how the W1 magnitude can provide an indication of how distant the host galaxy is, noting that mid-infrared faintness has shown to be an effective criterion for identifying high-redshift radio galaxies \citep[e.g.][]{Norris2006,Orenstein2019}. In Figure~\ref{fig:W1_by_medianspecz} we present the median spectroscopic redshift as a function of W1 magnitude, where we have only considered magnitude bins that contain a minimum of five spectroscopically-confirmed G4Jy sources. As expected, the median redshift shows an upwards trend as the W1 magnitude gets fainter, with the $z=1$ threshold being crossed at $\sim$15.25\,mag. We note that 20 per cent of the 3CRR sample (i.e. 35 sources) are at $z > 1$, whilst the same can be said for 5 per cent of the G4Jy Sample (i.e. 84 sources), bearing in mind that the latter (currently) has a spectroscopic completeness of 34 per cent \citep{White2026}. It is also worthwhile noting that the highest redshift in the sample -- acquired via NED, $z_{\mathrm{NED}} = 3.5699$ \citep{vanOjik1996} -- is for G4Jy 1020, with W1 = 17.638\,mag. Only eight other sources have a fainter W1 magnitude, and one of these is G4Jy 1456 (aka 4C 13.66) at $z_{\mathrm{NED}}=1.45$, which was the last 3CRR source to be spectroscopically confirmed \citep{1996MNRAS.279L..13R}.

\setkeys{Gin}{draft}

\subsection{Near-infrared properties}
\label{sec:NIR_properties}

% Lectures by Richard Ellis:
% https://ned.ipac.caltech.edu/level5/Ellis4/Ellis5.html

The $K$-band traces old stellar populations in galaxies, with the $K$-band luminosity shown by \citet{Kauffmann1998} to be a good measure of the underlying stellar mass of the system. This holds out to high redshift, except in the case of quasars, where the near-infrared continuum is dominated by the accretion disc rather than host-galaxy light. As such, the $K$--$z$ relation can be used as a diagnostic for distinguishing quasars from the general galaxy population.

\begin{figure}
\centering
\subfigure[$K$--$z$ space, by redshift origin]{\includegraphics[scale=0.52]{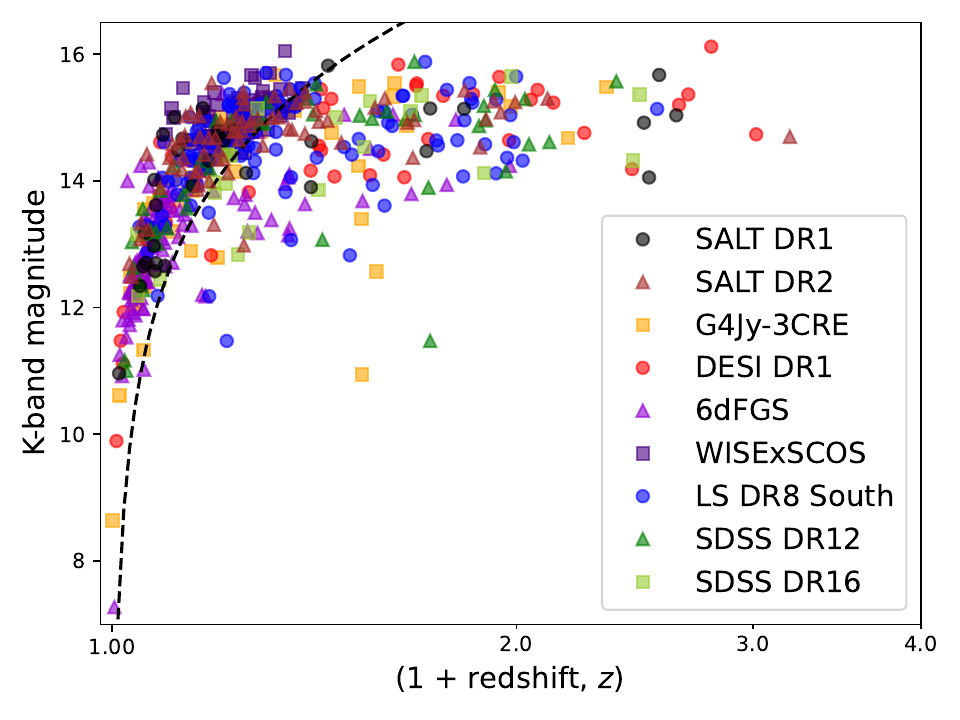}}
\subfigure[$K$--$z$ space, by spectral curvature]{\includegraphics[scale=0.52]{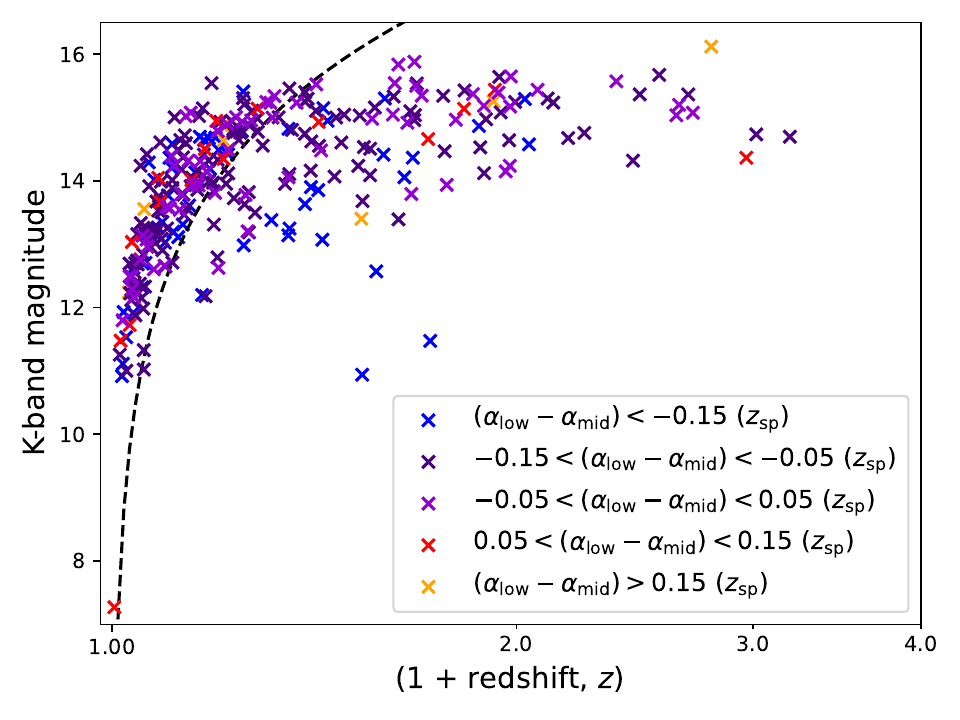}}
\subfigure[$K$--$z$ space, by optical morphology]{\includegraphics[scale=0.52]{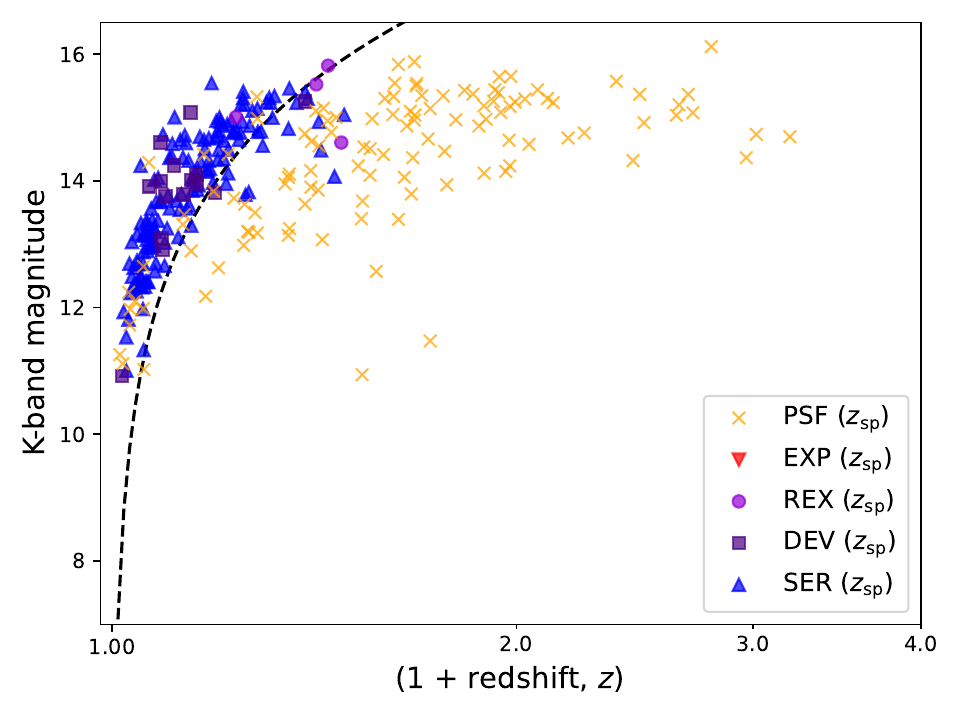}}
\caption{The $K$--$z$ datapoints for the G4Jy Sample (Section~\ref{sec:NIR_properties}), colour-coded by (a) the origin of the spectroscopic/photometric redshift, (b) the degree of spectral curvature in the radio (Equation~\ref{eq:sci0}), and (c) the optical morphology (Table~\ref{tab:sersic_index}). A reminder that `WISExSCOS' and `LS DR8 South' are photometric-redshift catalogues, and so they do not have markers that appear in panels (b) and (c), which involve purely spectroscopic redshifts ($z_{\mathrm{sp}}$). In each panel, the dashed line represents the empirical relation found by \citet{Willott2003} through $K$-band imaging for radio galaxies in the 7C Redshift Survey \citep[7CRS;][]{Willott1998} and other samples (incl. 3CRR).}
\label{fig:K_z_relation}
\end{figure}

In Figure~\ref{fig:K_z_relation} we plot $K$--$z$ datapoints for the G4Jy Sample, firstly as a function of the origin of the redshift measurement (Figure~\ref{fig:K_z_relation}a). This is mainly for interest, as we do not expect to see any particular trends with respect to the different redshift origins. Despite 6dFGS \citep{Jones2009} being a relatively-shallow optical survey, its datapoints span a large area of the parameter space, up to $z \sim 1$, and (relevant for the y-axis) we remind the reader that the 10-$\sigma$ depth of the $K_{S}$-band data from 2MASS is 14.3\,mag (Section~\ref{sec:NIR_data}). Also plotted (in all three panels of Figure~\ref{fig:K_z_relation}) is the empirical relation found by \citet{Willott2003} for radio galaxies in the 7C Redshift Survey \citep{Willott1998} and other Cambridge-based surveys (e.g. 3CRR). This was derived by fitting $K$-band magnitudes obtained through a 64-kpc aperture (so taking the redshift of the source into account). Interestingly, they found a remarkable correspondence in this fitted $K$--$z$ relation with the ``apparent $K$-magnitude evolution of a galaxy of local luminosity $3 L_*$ [that] forms all of its stars in an instantaneous burst at $z_{\mathrm{f}}=10$''. We plan to follow up on this when we have star-formation histories for the G4Jy Sample.

\setkeys{Gin}{draft=false}

\setkeys{Gin}{draft}

Next, in Figure~\ref{fig:K_z_relation}b, we plot $K$--$z$ parameter space as a function of spectral curvature, SCI$_0$ (Equation~\ref{eq:sci0}), for G4Jy-NVSS sources. This is in order to see whether there is any correspondence between the (radio) spectral behaviour of the source and its $K_{S}$ magnitude (a proxy for stellar mass) or redshift. As we saw in Figure~\ref{fig:colourcoded_WISE}b, there appears to be no correspondence, again suggesting a disconnect between the AGN lifecycle and the host galaxy. {\bf This points towards an external factor that governs the AGN lifecycle, such as the frequency with which gas may lose sufficient angular momentum in order to accrete onto the supermassive black-hole.} The magnetic field within the environment may also play an important role, as discussed for Figure~\ref{fig:PDdiagrams}b. 

Lastly, in Figure~\ref{fig:K_z_relation}c we present $K$--$z$ space as a function of optical morphology (as classified for the Legacy Surveys DR10; \citealt{Dey2019}). As expected, there is a predominance of sources with PSF-fitted light-profiles (i.e., quasars) to the right of the $K$--$z$ relation determined by \citet{Willott2003}. (For reference, 20 per cent of the G4Jy Sample has a host galaxy described by a `PSF' light-profile.) A few sources of other optical morphology can also be found to the right of the dashed line, and we suggest that these are starbursts. This is because the connection between stellar mass and the apparent $K$-band magnitude no longer holds for extreme starbursting systems \citep[e.g.][]{Kauffmann1998}, with the additional possibility that the starburst is combined with a quasar as part of a gas-rich merger \citep[e.g.][]{Hopkins2008}. Furthermore, we confirm that there are no sources with exponential-fitted light-profiles (`EXP'; i.e., spiral galaxies) that have $K$-band magnitudes. This may be because their stellar masses are too low for them to be detected in 2MASS, and/or they are extremely dust-obscured \citep[][]{daCunha2010,Triani2021}.

\section{Conclusions}
\label{sec:conclusions}

The previous paper in the series, Paper III, detailed the additional work that has gone into host-galaxy identification for the G4Jy Sample, alongside careful assessment of the various redshifts that are available in the literature. The resulting catalogue, which is further updated during the current work, Paper IV (Section~\ref{sec:catalogue}), forms the basis of our in-depth multiwavelength analysis of G4Jy sources, where we consider their radio, mid-infrared and near-infrared properties (Section~\ref{sec:results}). A unique feature of this analysis is the incorporation of radio-spectral-curvature information, and we summarise our conclusions as follows:

\begin{enumerate}
 %   \item{ [Cross-checks and mistaken identifications] }
    \item{ Crossmatching the G4Jy Sample with DR10 of the (DESI) Legacy Surveys \citep[LS;][]{Dey2019} results in $griz$ photometry for 1,299 sources. These surveys also provide optical morphology in the form of light-profile fitting by {\sc The Tractor} \citep{Lang2016}. This information, in addition to basic ancillary data from the NASA/IPAC Extragalactic Database, is provided in the multiwavelength G4Jy catalogue (Section~\ref{sec:catalogue}, Appendix~\ref{app:catalogue}). }
    \item{ We leverage the excellent spectral coverage provided by the MWA (over 72--231\,MHz) to study the degree of (radio) spectral curvature across the sample. This is simplified through the comparison of spectral indices ($\alpha$, following $S_{\nu} \propto \nu^{\alpha}$), where the spectral curvature index (SCI$_0$) is defined as $\alpha_{\mathrm{low}} - \alpha_{\mathrm{mid}}$. (For our study, $\alpha_{\mathrm{low}} = \alpha_{\mathrm{72\,MHz}}^{\mathrm{231\,MHz}}$ and $\alpha_{\mathrm{mid}} = \alpha_{\mathrm{151\,MHz}}^{\mathrm{1400\,MHz}}$.) After dividing the G4Jy Sample into five spectral-curvature bins, we propose that sources with SCI$_0 < -0.15$ are candidate restarted radio-galaxies, whilst sources with SCI$_0 >0.15$ are candidate remnant radio-galaxies. (This will be tested through visual inspection of deeper, higher-resolution radio-images than those currently available, in addition to modelling the age of the radio emission.) We note that observational bias may explain the tendency for {\it spectroscopically-confirmed} candidate remnant radio-galaxies to be found at lower redshift than the rest of the radio-galaxy population. Moreover, spectral flattening at low frequencies (due to absorption effects) complicates the interpretation of the SCI parameter as a proxy for spectral age. This will be explored in future work. }
    \item{ No trend is seen in spectral curvature whether the 151-MHz radio-luminosities or the 1400-MHz radio-luminosities are presented. We also present the linear sizes as a function of spectral curvature, but no trends are evident either.} 
    \item{ Following the 1-Mpc threshold of \citet{Willis1978}, we list 11 GRGs in Table~\ref{tab:GRGs}, noting that the angular sizes provided there are larger than those in the G4Jy catalogue, which need to be refined. }
    \item{ {\bf The radio-power--size ($P$--$D$) diagram \citep{Baldwin1982} is an informative way of studying the evolutionary paths of radio galaxies}, where sources move from the left-hand side to the right-hand side of the diagram as they expand and fade during their lifetime \citep[see also][]{Ryle1967}. Following on from \citet{White2025a}, we present the $P$--$D$ diagram as a function of radio morphology, and note the distinctive distribution of `single' sources (Figure~\ref{fig:PDdiagram}). Those that belong to a trunk-like region in $P$--$D$ space likely have radio-jets that are significantly slowed by the surrounding medium (and so are `frustrated' radio-sources), or are simply {\it young} radio-sources. Meanwhile, the `single' sources that span a range of linear sizes beyond this `trunk' are interpreted as FR-I or FR-II radio-galaxies that cannot be spatially resolved in NVSS imaging. Also, as expected, the `doubles' and `triples' extend to larger linear sizes for a range of radio luminosities \citep[see also the distribution of FR-II radio-galaxies in figure 6 by][]{Chilufya2025}.} 
    \item{ {\bf For the first time, we combine the radio luminosity and the linear size with the degree of spectral curvature, i.e. [$P$--$D$](SCI$_0$), which approximates the spectral age of the source (Figure~\ref{fig:PDdiagrams}b)} -- although, see the caveat above [item (ii)]. This leads to two intriguing observations: (a) that there is a predominance of candidate remnant radio-galaxies at $D<200$\,kpc (suggesting that the radio-jets are growing slowly on the same spatial scales as the circumgalactic medium), and (b) that candidate restarted radio-galaxies extend to GRG sizes (suggesting that the supermassive black-hole has been reignited following a long `duty cycle' of the AGN). However, the former interpretation is complicated by the presence of young radio-sources, which exhibit similar SCI$_0$ values as remnant radio-galaxies. }
    \item{ As previously shown by \citet{Gurkan2014, Jarrett2017}, {\color{black}and other authors}, radio galaxies populate the entirety of  {\it WISE} colour-colour space. We see no trend with respect to either the radio morphology or the spectral curvature of the G4Jy source, but there is correspondence between the sources whose (optical) light-profile is PSF-fitted and the `quasar' region of this colour-colour space. }
    \item{Investigation of the W1-magnitude distribution appears to confirm that the `excess' of bright magnitudes (with respect to the distrbution for Radio Galaxy Zoo; \citealt{Wong2025}) could be explained by low-redshift ($z < 0.5$) G4Jy sources that have diffuse radio-emission and are harboured by clusters. In addition, we find that the median spectroscopic-redshift crosses the $z=1$ threshold at $W1\sim15.25$\,mag, which demonstrates the effectiveness of using mid-infrared faintness to identify candidate high-redshift radio-galaxies. }
    \item{ The $K$-band magnitude is a good proxy for the stellar mass of the host galaxy, with the well-studied $K$--$z$ relation \citep{Willott2003} allowing for quasars to be distinguished from the rest of the galaxy population. We see that this is the case for G4Jy sources when the optical morphology (`PSF') is considered as a quasar classification. No trends are seen when datapoints are plotted in $K$--$z$ space as a function of redshift origin or spectral curvature of the radio emission. }
    \item{ To repeat: {\bf no trends are seen with respect to spectral curvature and the location of the source in either {\it WISE} colour-colour space or $K$--$z$ space. This suggests an expected disconnect between spectral-curvature properties and the host galaxy, which points towards an external factor that influences the radio lifecycle of the source.} Given the recurrent activity that we witness for restarted radio-galaxies, this could be the frequency with which gas accretes onto the central supermassive black-hole. }
\end{enumerate}

\vspace{5mm}
As shown, {\it multiwavelength} data analysis is crucial for investigating the physical processes that govern how galaxies evolve. The  {\it legacy multiwavelength dataset} that we have compiled in this work (Papers III and IV; White et al., 2025c, 2025d) will therefore enable detailed analysis of the G4Jy Sample, and we expect this to be supplemented by many more datasets, such as X-ray information.

\section*{Acknowledgements}

We thank the referee for their comments, which helped to improve the manuscript. In addition, we thank Dustin Lang and Ross Turner for helpful discussions, and those involved in the Astro Data Lab Science Platform (\url{https://datalab.noirlab.edu/}).

This work is based on the research supported in part by the National Research Foundation of South Africa (Grant Number 151060). The financial assistance of the South African Radio Astronomy Observatory (SARAO) towards this research is also hereby acknowledged. CJR acknowledges support from the DFG via the Collaborative Research Center SFB1491, \textit{Cosmic Interacting Matters -- From Source to Signal} (project no. 445052434).

Some of the observations reported in this paper were obtained with the Southern African Large Telescope (SALT), under program 2020-1-MLT-008 (PI: White). Additional data were obtained via the MeerKAT telescope (via proposal SCI-20190418-SW-01). This is operated by SARAO, which is a facility of the National Research Foundation, an agency of the Department of Science and Innovation.

This research uses services or data provided by the Astro Data Lab, which is part of the Community Science and Data Center (CSDC) Program of NSF NOIRLab. NOIRLab is operated by the Association of Universities for Research in Astronomy (AURA), Inc. under a cooperative agreement with the U.S. National Science Foundation.

This research has made use of the NASA/IPAC Extragalactic Database (NED), which is funded by the National Aeronautics and Space Administration and operated by the California Institute of Technology.

The National Radio Astronomy Observatory is a facility of the National Science Foundation operated under cooperative agreement by Associated Universities, Inc.

This publication makes use of data products from the Two Micron All Sky Survey, which is a joint project of the University of Massachusetts and the Infrared Processing and Analysis Center/California Institute of Technology, funded by the National Aeronautics and Space Administration and the National Science Foundation. 

This scientific work uses data obtained from Inyarrimanha Ilgari Bundara / the Murchison Radio-astronomy Observatory. We acknowledge the Wajarri Yamaji People as the Traditional Owners and native title holders of the Observatory site. CSIRO’s ASKAP radio telescope is part of the Australia Telescope National Facility (https://ror.org/05qajvd42). Operation of ASKAP is funded by the Australian Government with support from the National Collaborative Research Infrastructure Strategy. ASKAP uses the resources of the Pawsey Supercomputing Research Centre. Establishment of ASKAP, Inyarrimanha Ilgari Bundara, the CSIRO Murchison Radio-astronomy Observatory and the Pawsey Supercomputing Research Centre are initiatives of the Australian Government, with support from the Government of Western Australia and the Science and Industry Endowment Fund. This paper includes archived data obtained through the CSIRO ASKAP Science Data Archive, CASDA (https://data.csiro.au).

Funding for the Sloan Digital Sky Survey IV has been provided by the Alfred P. Sloan Foundation, the U.S. Department of Energy Office of Science, and the Participating Institutions. SDSS-IV acknowledges support and resources from the Center for High Performance Computing  at the University of Utah. The SDSS website is www.sdss.org. SDSS-IV is managed by the Astrophysical Research Consortium for the Participating Institutions of the SDSS Collaboration including the Brazilian Participation Group, the Carnegie Institution for Science, Carnegie Mellon University, Center for Astrophysics | Harvard \& Smithsonian, the Chilean Participation Group, the French Participation Group, Instituto de Astrof\'isica de Canarias, The Johns Hopkins University, Kavli Institute for the Physics and Mathematics of the Universe (IPMU) / University of Tokyo, the Korean Participation Group, Lawrence Berkeley National Laboratory, Leibniz Institut f\"ur Astrophysik Potsdam (AIP), Max-Planck-Institut f\"ur Astronomie (MPIA Heidelberg), Max-Planck-Institut f\"ur Astrophysik (MPA Garching), Max-Planck-Institut f\"ur Extraterrestrische Physik (MPE), National Astronomical Observatories of China, New Mexico State University, New York University, University of Notre Dame, Observat\'ario Nacional / MCTI, The Ohio State University, Pennsylvania State University, Shanghai Astronomical Observatory, United Kingdom Participation Group, Universidad Nacional Aut\'onoma de M\'exico, University of Arizona, University of Colorado Boulder, University of Oxford, University of Portsmouth, University of Utah, University of Virginia, University of Washington, University of Wisconsin, Vanderbilt University, and Yale University.

%%%%%%%%%%%%%%%%%%%%%%%%%%%%%%%%%%%%%%%%%%%%%%%%%%
%\vspace{-7mm}

\section*{Data Availability}

The updated G4Jy catalogue will be made available through VizieR \citep{Ochsenbein2000} and the GitHub repository for the G4Jy Sample: \url{https://github.com/svw26/G4Jy}. Overlays for the full sample, amongst other G4Jy data, can be downloaded from the Zenodo repository: \url{https://zenodo.org/communities/g4jy/}. We encourage others to also make their G4Jy data available here, in the interest of Open Access, discoverability, and general best practices \citep{Chen2022}. 

%\vspace{-7mm}



\bibliographystyle{mnras}
\bibliography{G4Jy_Paper_IV}

%\begin{thebibliography}{99}
%\end{thebibliography}

%%%%%%%%%%%%%%%%%%%%%%%%%%%%%%%%%%%%%%%%%%%%%%%%%%

%%%%%%%%%%%%%%%%% APPENDICES %%%%%%%%%%%%%%%%%%%%%

\appendix

\section{Columns (and an example row) of the multiwavelength G4Jy catalogue}
\label{app:catalogue}

Descriptions of newly-added/newly-named columns in the multiwavelength G4Jy catalogue, along with ninth-row entries, can be found in Table~\ref{tab:catalogue_columns}. Readers who wish to combine the current catalogue with the original G4Jy catalogue \citep[][]{White2020a, White2020b}, in order to retrieve all of the low-frequency radio data (see appendix E of Paper I), are advised to crossmatch the catalogues on the centroid positions (centroid\_RAJ2000, centroid\_DEJ2000). This is because, unlike the host-galaxy positions, the centroid positions exist for the full sample. In addition, the latter remain fixed, which helps with long-term `future-proofing' of the catalogue. 

%\begin{landscape}
\begin{table*}
\centering
\caption{Column numbers, names, units, descriptions, and ninth-row entries for 60 of the 131 columns in the updated G4Jy catalogue (Section~\ref{sec:catalogue}, Appendix~\ref{app:catalogue}). This row has been chosen to minimise the number of empty columns. Column names with the suffix `LSDR10' indicate that the column data has been obtained from the Legacy Surveys DR10 \citep{Dey2019}. Similarly, `wiseScos' refers to the photometric-redshift catalogue of \citet{Bilicki2016}, and `LSDR8south' refers to the photometric-redshift catalogue of \citet{Duncan2022}. See appendix~E of \citet{White2020b} for the remaining 71 G4Jy-catalogue columns.}
\begin{tabular}{@{}ccccr@{}} 
\hline
Column no. & Column name & Units & Description & Ninth-row entry  \\
\hline
2 & NED\_name & -- & Alternative name for the G4Jy source in NED & PKS 0003$-$56 \\
21 & S\_1400\_Jy & Jy & Total flux-density at 1400 MHz (extrapolated for SUMSS sources) &  2.1010 \\
22 & err\_S\_1400\_Jy & Jy & Error on the total flux-density at 1400 MHz &  0.1413 \\
23 & alpha\_mid & -- & Spectral index, 151 to 1400 MHz (extrapolated for SUMSS sources) &  $-$0.678867 \\
24 & err\_alpha\_mid & -- & Error on the spectral index between 151 and 1400 MHz &  0.030214 \\

69 & host\_name & -- & Name of the host galaxy   &  WISEA J000557.94$-$562831.0 \\
70 & host\_RAJ2000 & deg & Right ascension for the host galaxy (J2000)  & 1.4914452 \\
71 & host\_DEJ2000 & deg & Declination for the host galaxy (J2000)  & $-$56.475300 \\
80 & eeMaj\_allwise & arcsec & Semi-major axis of the error ellipse  & 0.0508 \\
81 & eeMin\_allwise & arcsec & Semi-minor axis of the error ellipse  & 0.0500 \\
82 & eePA\_allwise & deg & Position angle of the error ellipse  & 77.7 \\
83 & Jmag\_allwise & mag & $J$ magnitude from 2MASS via the AllWISE catalogue  & 16.343 \\
84 & err\_Jmag\_allwise & mag & Error on $J$ magnitude from 2MASS  & 0.158 \\
85 & Hmag\_allwise & mag & $H$ magnitude from 2MASS via the AllWISE catalogue  & 15.493 \\
86 & err\_Hmag\_allwise & mag & Error on $H$ magnitude from 2MASS  & 0.188 \\
87 & Kmag\_allwise & mag & $K_{S}$ magnitude from 2MASS via the AllWISE catalogue  & 14.781 \\
88 & err\_Kmag\_allwise & mag & Error on $K_{S}$ magnitude from 2MASS  & 0.122 \\
89 & ra\_LSDR10 & deg & Right ascension in Legacy Surveys DR10 (J2000)  & 1.491361150872304 \\
90 & dec\_LSDR10 & deg & Declination in Legacy Surveys DR10 (J2000)  & $-$56.47533820865902 \\
91 & mag\_g\_LSDR10 & mag & $g$ magnitude in Legacy Surveys DR10  & 18.377111 \\
92 & mag\_r\_LSDR10 & mag & $r$ magnitude in Legacy Surveys DR10  & 16.903824 \\
93 & mag\_i\_LSDR10 & mag & $i$ magnitude in Legacy Surveys DR10  & 16.386793 \\
94 & mag\_z\_LSDR10 & mag & $z$ magnitude in Legacy Surveys DR10  & 16.092787 \\
95 & dered\_mag\_g\_LSDR10 & mag & Dereddenned $g$ magnitude in Legacy Surveys DR10  & 18.342104 \\
96 & dered\_mag\_r\_LSDR10 & mag & Dereddenned $r$ magnitude in Legacy Surveys DR10  & 16.880241 \\
97 & dered\_mag\_i\_LSDR10 & mag & Dereddenned $i$ magnitude in Legacy Surveys DR10  & 16.369452 \\
98 & dered\_mag\_z\_LSDR10 & mag & Dereddenned $z$ magnitude in Legacy Surveys DR10  & 16.079596 \\
99 & snr\_g\_LSDR10 & -- & Signal-to-noise ratio in the $g$ band  & 536.82965 \\
100 & snr\_r\_LSDR10 & -- & Signal-to-noise ratio in the $r$ band  & 1232.6284 \\
101 & snr\_i\_LSDR10 & -- & Signal-to-noise ratio in the $i$ band  & 1319.4358 \\
102 & snr\_z\_LSDR10 & -- & Signal-to-noise ratio in the $z$ band  & 879.1196 \\
103 & psfsize\_g\_LSDR10 & arcsec & Weighted average PSF FWHM in the $g$ band  & 1.3799053 \\
104 & psfsize\_r\_LSDR10 & arcsec & Weighted average PSF FWHM in the $r$ band  & 1.0936546 \\
105 & psfsize\_i\_LSDR10 & arcsec & Weighted average PSF FWHM in the $i$ band  & 1.0151794 \\
106 & psfsize\_z\_LSDR10 & arcsec & Weighted average PSF FWHM in the $z$ band  & 1.0915892 \\
107 & morphtype\_LSDR10 & -- & Optical morphological classification by {\sc The Tractor}  & SER \\
108 & sersic\_LSDR10 & -- & S{\' e}rsic-index value fitted/set by {\sc The Tractor}  & 5.3008156 \\
109 & L\_151\_WperHz & W\,Hz$^{-1}$ & K-corrected radio luminosity at 151 MHz  &  2.430700E27 \\
110 & err\_L\_151\_WperHz & W\,Hz$^{-1}$ & Error on the K-corrected radio luminosity at 151 MHz  & 1.553722E25 \\
111 & L\_1400\_WperHz & W\,Hz$^{-1}$ & K-corrected radio luminosity at 1400 MHz &  5.204895E26 \\
112 & err\_L\_1400\_WperHz & W\,Hz$^{-1}$ & Error on the K-corrected radio luminosity at 1400 MHz &  3.931781E25 \\
113 & linear\_size\_limit & -- & Indicates that the linear size is an upper limit &  -- \\
114 & linear\_size\_kpc & kpc & Linear size (via the angular size in NVSS/SUMSS)  & 249.0065 \\

115 & z\_origin\_flag & -- & Flag denoting the origin of the adopted redshift &  3.0 \\
116 & z\_Quality\_flag & -- & Flag indicating the quality of the adopted redshift (1 = best) &  1 \\
117 & zsp\_misc & -- & Miscellaneous spectroscopic redshift &  0.291 \\
118 & zsp\_DESI & -- & Spectroscopic redshift fitted by the DESI collaboration &  -- \\
119 & zsp\_WARN\_DESI & -- & Redshift-fitting quality flag by the DESI collaboration &  -- \\
120 & zph\_misc & -- & Miscellaneous photometric redshift &  0.279 \\
121 & zph\_LSDR8south & -- & Photometric-redshift estimate from LS DR8 South &  0.279 \\
122 & err\_zph\_LSDR8south & -- & Error in the photometric-redshift estimate &  0.027 \\
123 & zph\_fclean\_LSDR8south & -- & Flag = 1 for sources free of blending or imaging artefacts &  1 \\
124 & zph\_fqual\_LSDR8south & -- & Flag = 1 for reliable photometric-redshift estimates &  1 \\
125 & zph\_ANN\_wiseScos & -- & Photometric redshift obtained with the ANNz framework &  0.269903 \\
126 & ebv\_wiseScos & mag & E(B-V), the Galactic dust extinction along the line-of-sight &  0.011 \\
127 & NED\_RAJ2000 & deg & Fiducial right ascension in NED (J2000) &  1.491445 \\
128 & NED\_DEJ2000 & deg & Fiducial declination in NED (J2000) &  $-$56.4753 \\
129 & NED\_type & -- & Object type accompanying the alternative name in NED &  G \\
130 & z\_NED & -- & Fiducial redshift (spectroscopic or photometric) in NED &  0.2912 \\
131 & z\_ref\_NED & -- & Reference for the fiducial redshift in NED &  20032dF...C...0000C \\

\hline
\label{tab:catalogue_columns}
\end{tabular}
\end{table*}
%\end{landscape}

%\setcounter{table}{0} 

\section{More on spectral indices}
\label{app:alpha}

\setkeys{Gin}{draft=false}

\begin{figure}
\centering
\subfigure[]{
\includegraphics[scale=0.51]{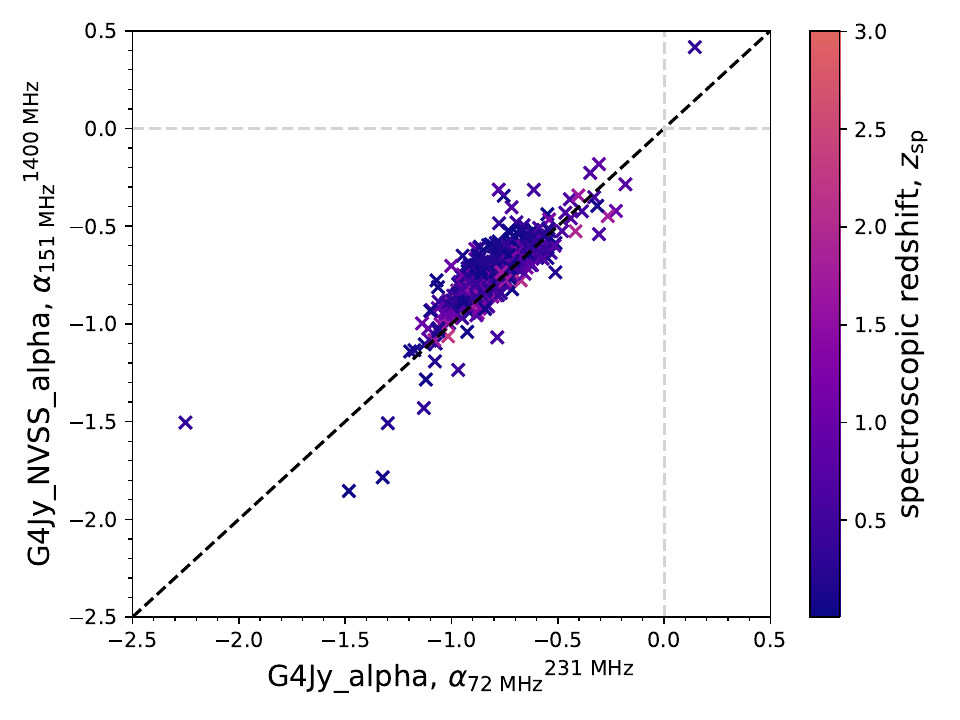} }
\subfigure[]{
\includegraphics[scale=0.51]{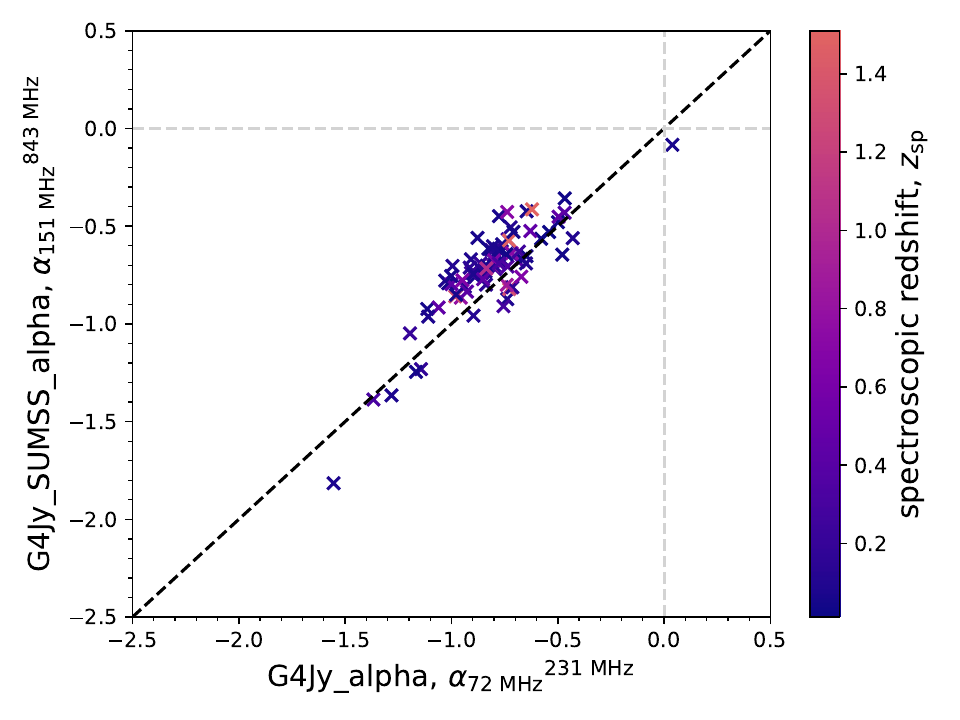} }
\caption{Panels (a) and (b) show different spectral indices from the G4Jy catalogue \citep{White2020a,White2020b} plotted against each other (Section~\ref{sec:spectral_curvature}). Diagonal lines represent the 1:1 relation, which signifies the (narrow) region over which the radio emission between low and `mid' frequencies can accurately be described by a power-law function. The datapoints of each panel are colour-coded by the spectroscopic redshift to help look for any trend with spectral index (Appendix~\ref{app:alpha}). }
\label{fig:alpha_alpha_by_specz}
\end{figure}

Continuing the investigation of spectral curvature detailed in Section~\ref{sec:spectral_curvature}, we study the distribution of spectroscopic redshift within the $\alpha_{\mathrm{low}}-\alpha_{\mathrm{mid}}$ plane (Figure~\ref{fig:alpha_alpha_by_specz}). We see no tendency for the highest redshifts to be associated with the steepest spectral-indices, which calls into question the efficacy of common methods for identifying high-redshift radio-sources \citep[see also][]{Pinjarkar2025}. However, we note that a much smaller fraction of sources lies below the 1:1 relation than seen in Figure~\ref{fig:alpha_alpha}. This could be because more-recent AGN activity (relevant for datapoints above the 1:1 relation) correlates with greater photoionisation of the broad- and narrow-line regions, making spectroscopic redshifts easier to obtain. This is in contrast to ageing sources (relevant to the datapoints below the 1:1 relation), which may have little to no AGN-related optical emission.

\begin{figure*}
\centering
\subfigure[]{
\includegraphics[scale=0.85]{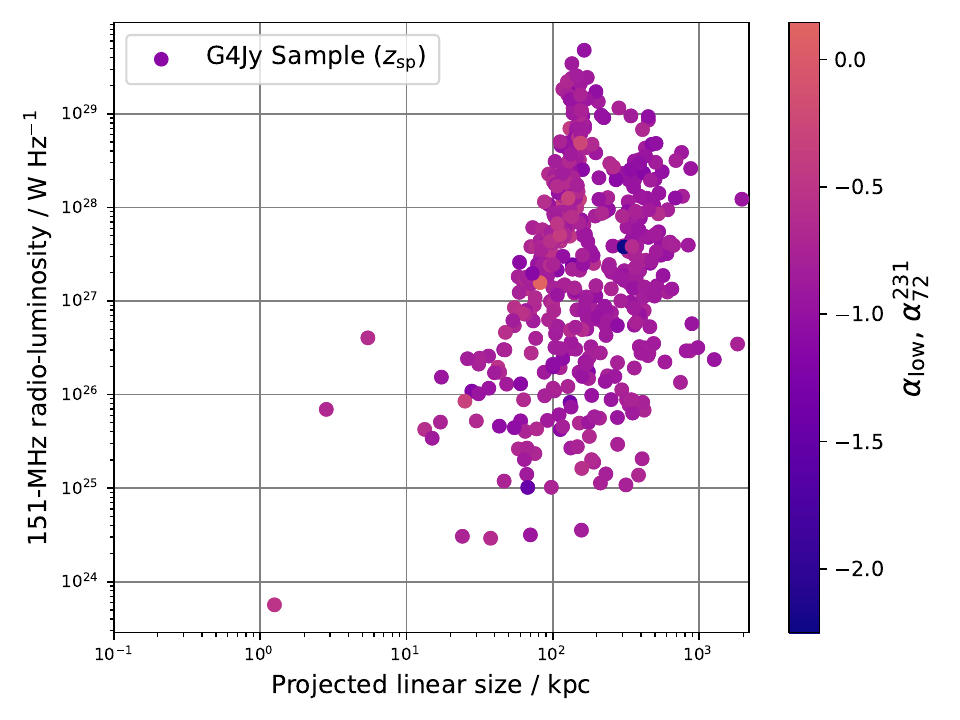} }
\subfigure[]{
\includegraphics[scale=0.85]{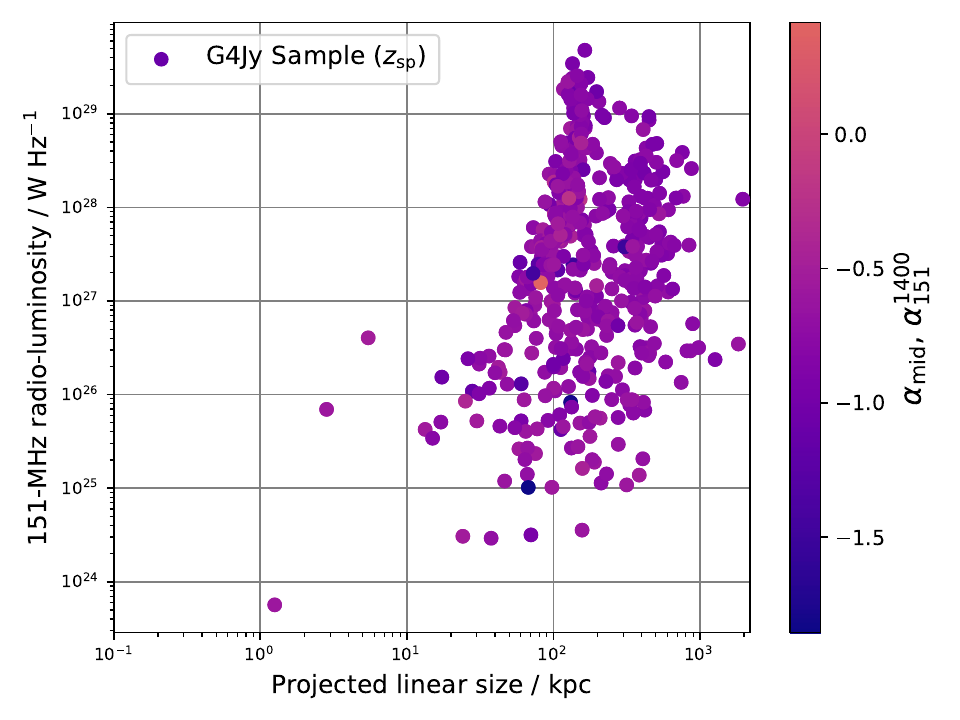} }
\caption{We present the $P$--$D$ diagram (Section~\ref{sec:PDdiagrams}) as a function of (a) $\alpha_{\mathrm{low}}$ and (b) $\alpha_{\mathrm{mid}}$ for G4Jy-NVSS sources. These spectral indices correspond to the G4Jy\_alpha and alpha\_mid (i.e. G4Jy\_NVSS\_alpha) values in the updated G4Jy catalogue, respectively. As discussed in Appendix~\ref{app:alpha}, neither spectral index appears to demarcate any particular region of $P$--$D$ space. }
\label{fig:PD_by_alpha}
\end{figure*}

We also check whether any trends can be seen if the $P$--$D$ diagram (Section~\ref{sec:PDdiagrams}) is plotted as a function of $\alpha_{\mathrm{low}}$ or $\alpha_{\mathrm{mid}}$. As shown in Figure~\ref{fig:PD_by_alpha}, and as expected, the distributions appear `smooth' with respect to the spectral index that is used. This is because a single spectral index can take on a wide range values that are all consistent with synchrotron emission. Clearly, spectral {\it curvature} is key to exploring the evolutionary states of radio galaxies in further detail (Figure~\ref{fig:PDdiagrams}b), and we look forward to having more-quantitative results as the G4Jy Sample approaches spectroscopic completeness.

\setkeys{Gin}{draft}

\section{Giant radio-galaxies}
\label{sec:GRGs}

In Table~\ref{tab:GRGs} we list the sources that pass the 1-Mpc threshold for categorisation as giant radio-galaxies (GRGs), following the definition by \citet[][]{Willis1978}. Three of these (G4Jy 318, G4Jy 1021, and G4Jy 1079) have new angular-size measurements via VLASS, and we describe another two sources in detail below. In addition, a further three G4Jy sources warrant special discussion in the text. We also note that, as part of creating Table~\ref{tab:GRGs}, we discard three `GRGs' (G4Jy 155, G4Jy 285, and G4Jy 315) where the linear size is determined via a photometric redshift that is flagged by \citet{Duncan2022} as being `unreliable'.

%\begin{landscape}
\begin{table*}%[]
    \centering
    \begin{tabular}{c|c|c|c|c|c|c|c}
    \hline
    %\multicolumn{2}{|c|}{\text{Host-galaxy}}
    Source name & Angular & Redshift & Redshift, & Redshift & z\_Quality\_flag & Linear  & Reported \\ 
    & size / arcmin & type & $z$ & reference & (Paper III) & size / kpc  
     & as a GRG by \\ 
       \hline

%G4Jy 79 & 6.48 & $z_{\mathrm{ph}}$ & 0.161   & \citet{Duncan2022}  & 5 & 1078    \\ % spec-z shows that the linear size is too small
        G4Jy 133 & 11.80 & $z_{\mathrm{sp}}$ & 0.1478 & \citet{GarciaPerez2024} & 1 & 1828 & \citet{Andernach2021}  \\ % 2.582 kpc/"
  %   G4Jy 155 & 2.44 & $z_{\mathrm{ph}}$ &  1.437  & \citet{Duncan2022} & 5 & 1239    \\ % Unreliable photo-z
     G4Jy 234 & 6.30 & $z_{\mathrm{ph}}$ & 0.189    & \citet{Duncan2022}  & 5 & 1193  & \citet{Andernach2021} \\ % 3.156 kpc/"
   %     G4Jy 285 & 9.64 & $z_{\mathrm{ph}}$ & 0.110 & \citet{Duncan2022} & 5 & 1161   \\ % Unreliable photo-z
  %   G4Jy 315 & 7.35 & $z_{\mathrm{ph}}$ &  0.152  & \citet{Duncan2022} & 5 & 1165    \\ % Unreliable photo-z
G4Jy 318 & 3.56 & $z_{\mathrm{ph}}$ & 0.736   & \citet{Duncan2022}  & 5 & 1556 & This work \\ % 7.286 kpc/"
        G4Jy 347 & 25.86 & $z_{\mathrm{sp}}$ & 0.0624 & \citet{GarciaPerez2024} & 1 & 1865  & \citet{Andernach2021} \\ % 1.202 kpc/"
     G4Jy 517 & 38.50 & $z_{\mathrm{sp}}$ &  0.03812  & \citet{Jones2009} & 1 & 1746  &  \citet{Saripalli1986} \\ % 0.756 kpc/"

G4Jy 641 & 7.77 & $z_{\mathrm{sp}}$ & 0.128 & \citet{White2025a} & 1 & 1070 & This work \\

     G4Jy 923 & 4.90 & $z_{\mathrm{sp}}$ & 0.6337    & (re-fitted, Paper III)  & 1 & 2010  & \citet{Bhatnager1998} \\ % 6.852 kpc/"
     G4Jy 1021 & 2.80 & $z_{\mathrm{ph}}$ & 0.931 & \citet{Duncan2022} & 5 & 1321 & This work \\ % 7.864 kpc/"
        G4Jy 1079 & 14.90 & $z_{\mathrm{sp}}$ &  0.0836  & \citet{GarciaPerez2024} & 1 & 1404  & \citet{Machalski2007} \\ % 1.571 kpc/"
             G4Jy 1282 & 19.50 & $z_{\mathrm{sp}}$ &  0.08745 & \citet{AbdulKarim2025} & 1 & 1914 & \citet{Willis1978} \\ % aka 3C 326, 1.636 kpc/"
            % G4Jy 1289?? & & \\ % Haven't confirmed the core yet
     G4Jy 1525 & 6.10 & $z_{\mathrm{NED}}$ &  0.3462  & \citet{2005AJ....130..896S} & 5 & 1795  & \citet{Subrahmanyan1996}  \\ % aka PKS 1910-80. 4.905 kpc/" 

   \hline
  \end{tabular}
    \caption{Angular sizes, redshifts, and (projected) linear sizes for 11 GRGs that are larger than 1\,Mpc (Section~\ref{sec:linearsizes}). We note that angular sizes catalogued by \citet{White2020a,White2020b} are based on NVSS and SUMSS imaging, which we revise here, based on the most-recent, available
surveys. `Redshift type' indicates whether the redshift is spectroscopic ($z_{\mathrm{sp}}$), photometric ($z_{\mathrm{ph}}$), or found through NED ($z_{\mathrm{NED}}$). A guide to the assigned z\_Quality\_flags is provided in Section~\ref{sec:z_information}. } 
    \label{tab:GRGs}
\end{table*}
%\end{landscape}
% Moreover, the asterisk (*) indicates angular sizes that have been derived more-recently, as described in the text. 

We first turn our attention to {\bf G4Jy 79}, which is quoted as having a linear size of 1078\,kpc in the G4Jy catalogue. However, this is based on a photometric redshift of $z_{\mathrm{ph}}=0.161$, whereas $z_{\mathrm{sp}}=0.123$ has been measured by \citet{Machalski2007}. Combined with an angular size of 5.9\,arcmin, as measured in VLASS, the linear size is 780\,kpc. Hence, it does not meet our 1-Mpc criterion for a GRG, and so we do not list it in Table~\ref{tab:GRGs}. 
%G4Jy 79 had been found by 2007AcA....57..227Machalski+ who  measured z\_sp=0.123 and is listed in 2017MNRAS.471.3806Kuzmicz+ as a possible DDRG, but RACS and VLASS show that there are just diffuse plumes, no inner FR II, so no DDRG.It is in the compilation by 2018ApJS..238....9Kuzmicz+ . With LAS=5.9' from VLASS, thus LLS=0.78 Mpc

Unfortunately, a photometric redshift in our catalogue leads to another inflated linear size (1165\,kpc), this time for {\bf G4Jy 315}. Also known as 3C 75, and previously presented in G4Jy Paper II \citep{White2020a}, this source is associated with a pair of wide-angle-tail radio-galaxies: NGC 1128N and NGC 1128S. They are 16\,arcsec apart from each other, and belong to cluster Abell 400 at a redshift of 0.0244 \citep[e.g.][]{Owen1995}. Having measured an angular size of 11.5\,arcmin in RACS-low \citep{McConnell2020}, we obtain 340\,kpc for its linear size and therefore conclude that G4Jy~315 (NGC 1128N) is not a GRG.   
%(or NGC 1128 NED01 and NGC 1128 NED02) . See 1995AJ....109...14Owen+ who did spectroscopy of both galaxies. Also 1999PASP..111..438Falco+.  Apparently the  N comp is at cz = 6780 km/s and the S comp around 7200 km/s, while the cluster mean is around 7300 km/s.

Several groups \citep{Tritton1973, Jones1992, Malarecki2013} have incorrectly proposed WISEA J070530.57$-$451311.1 as the host galaxy for {\bf G4Jy~641}, whereas the identification of \citet{Sejake2023}, WISEA J070532.94$-$451308.8, is supported by detecting the radio core in a MeerKAT image (see their figure 10). We use this image to measure the angular size (7.77\,arcmin) and, in combination with a SALT redshift ($z_{\mathrm{sp}}=0.128$; \citealt{White2025a}), we estimate the linear size as 1070\,kpc. We add this new GRG to Table~\ref{tab:GRGs} and note that the catalogued linear-size (585\,kpc) is based on an outdated angular-size of 4.26\,arcmin.   
% but your MeerKAT image clearly changes this. Well worth a note in paper III.
%I have 1973MNRAS.165..245Tritton+ for an early redshift %% SW: only see identifications, no redshifts
%  With LAS=7.77' from your MeerKAT image and z\_sp=0.128 (SALT)  I get LLS=1.07 Mpc.
%  The compact source just beyond the edge of the WSW lobe is separate with host SMSS J070517.33-451400.2,rP=18.43, WISEA J070517.35-451400.5:W1234=15.36,14.00,10.53,7.71, z\_GaiaDR3=0.08845
% Our host galaxy is WISEA J070532.94-451308.8. Heinz originally provided 2MASX J04590828$-$5252072 as the incorrect host
% Malarecki et al. (2013) quote (J2000) R.A. = 07:05:31, Dec. = $-$45:13:12

For {\bf G4Jy 923}, we can precisely measure the angular size as 4.90\,arcmin in VLASS. (This is only slightly above the catalogued angular-size of 4.77\,arcmin.) In addition, our re-fitted spectroscopic redshift (via DESI DR1) is in perfect agreement with $z_{\mathrm{sp}}= 0.6337$ from \citet{Bhatnager1998}. Thus, we calculate a linear size of 2010\,kpc (differing from the catalogued value of 1961\,kpc).
%precise LAS from VLASS is 4.9'  and z\_sp= 0.6337 from  1998MNRAS.299L..25B
%  on  " HE 1127-1304: the largest radio quasar" (which it is not any more ...). thus LLS=2.01 Mpc  

% namely as NVGRC 1106 by 2016ApJS..224...18Proctor

Finally, there has been considerable debate over the host galaxy of {\bf G4Jy 1289} (see section 3.2 of Paper III), which was first considered a candidate GRG by \citet{Proctor2016} through NVSS imaging. The MeerKAT image by \citet{Sejake2023} indicates that there are three candidates for the radio-core position: WISEA J155900.91$-$213935.5 \citep{White2026}, 2MASX J15590200$-$2140032 \citep{Quici2025}, and UGCS J155856.54$-$213824.3 \citep{Andernach2025}. Only when the radio core, and therefore the host-galaxy identification, has been confirmed, can we obtain the correct redshift for G4Jy 1289. This would then allow us to calculate the linear size and so determine whether or not this source is a GRG (in the sense of it passing the now, more-common threshold of 700\,kpc in projected linear-size).

%%%%%%%%%%%%%%%%%%%%%%%%%%%%%%%%%%%%%%%%%%%%%%%%%%

% Don't change these lines
\bsp	% typesetting comment
\label{lastpage}
\end{document}